\documentclass[usenatbib,useAMS]{mn2e}

\usepackage{natbib}
\usepackage{amsmath}
\usepackage{graphicx}
\usepackage{color}
\usepackage{epsfig}
\allowdisplaybreaks[1]

\newcommand{\msun}{{\rm M}_\odot}

\newcommand{\zsun}{Z_\odot}
\newcommand{\lsun}{{\rm L}_\odot}
\newcommand{\cc}{{\rm cm}^{-3}}
\newcommand{\mdot}{\dot m}
\newcommand{\msunyr}{{\rm M}_\odot~{\rm yr}^{-1}}

\newcommand{\K}{{\rm K}}
\newcommand{\beq}{\begin{equation}}
\newcommand{\eeq}{\end{equation}}

\title[Hyper-Eddington growth of IMBHs]
{Hyper-Eddington accretion flows onto massive black holes}

\author[]{Kohei Inayoshi$^{1}$\thanks{E-mail: inayoshi@astro.columbia.edu}
\thanks{Simons Society of Fellows, Junior Fellow},
Zolt$\acute{\rm a}$n Haiman$^{1}$,
Jeremiah P. Ostriker$^{1,2}$\\
$^{1}$Department of Astronomy, Columbia University, 550 W. 120th Street, New York, NY 10027, USA\\
$^{2}$Princeton University Observatory, Princeton, NJ 08544, USA}

\begin{document}
\maketitle
\label{firstpage}

\begin{abstract}
We study very-high rate, spherically symmetric accretion flows onto massive black holes 
(BH; $10^2 \la M_{\rm BH} \la 10^6~\msun$) embedded in dense metal-poor clouds, 
performing one-dimensional radiation hydrodynamical simulations.
We find solutions from outside the Bondi radius at hyper-Eddington rates,
unimpeded by radiation feedback when 
$(n_{\infty}/10^5~\cc) > (M_{\rm BH}/10^4~\msun)^{-1}(T_{\infty}/10^4~\K)^{3/2}$,
where $n_{\infty}$ and $T_\infty$ are the density and temperature of ambient gas.
Accretion rates in this regime are steady, and larger than $5000~L_{\rm Edd}/c^2$, 
where $L_{\rm Edd}$ is the Eddington luminosity.
At lower Bondi rates, the accretion is episodic due to radiative feedback 
and the average rate is below the Eddington rate.
In the hyper-Eddington case, the solution consists of 
a radiation-dominated central core, where photon trapping due to electron scattering is important, and 
an accreting envelope which follows a Bondi profile with $T\simeq 8000~\K$.
When the emergent luminosity is limited to $\la L_{\rm Edd}$ because of photon trapping,
radiation from the central region does not affect the gas dynamics at larger scales.
We apply our result to the rapid formation of massive BHs in protogalaxies with a virial temperature of 
$T_{\rm vir}\ga 10^4~\K$.
Once a seed BH forms at the center of the galaxy, it can grow to a maximum
$\sim 10^5~(T_{\rm vir}/10^4~\K)~\msun$ via gas accretion independent of the initial BH mass.
Finally, we discuss possible observational signatures of rapidly accreting BHs with/without allowance for dust.
We suggest that these systems could explain Ly$\alpha$ emitters without X-rays and
nearby luminous infrared sources with hot dust emission, respectively.
\end{abstract}



\section{Introduction}
\label{sec:intro}

Can very high accretion rates onto black holes (BHs) occur stably?
The existence of bright quasars (QSOs) at high-redshift ($z\ga 6$)
provides a challenging puzzle about the origin of supermassive black holes (SMBHs) 
with masses $\ga 10^9~\msun$
\citep{2006NewAR..50..665F, 2010AJ....139..906W, 2011Natur.474..616M, 
2013ApJ...779...24V,2014AJ....148...14B,2015Natur.518..512W}.
To form such massive objects within the first billion years after the Big Bang, 
rapid growth of seed BHs in the early Universe is required.

The initial seeds may originate via several scenarios
\citep[e.g.][references therein]{2012Sci...337..544V,2013ASSL..396..293H}.
One possibility is massive BHs formed through collapse of Population III (Pop III) stars
with typical mass of $\sim 100~\msun$
\citep{2008Sci...321..669Y, 2011Sci...334.1250H,2012MNRAS.422..290S,2014ApJ...781...60H,2015arXiv151001407H},
which grow up to SMBHs via continuous gas accretion or frequent major mergers of their host galaxies 
\citep{MadauRees01,HaimanLoeb01,VHM03,Li+07,2014Sci...345.1330A,2015ApJ...799..178K}.
A second possibility is the so-called direct collapse scenario, 
in which a supermassive star collapses by general relativistic instability and turns to a massive seed BH
with mass of $\ga 10^5~\msun$
\citep{1994ApJ...432...52L,2002ApJ...569..558O,2003ApJ...596...34B,
2006MNRAS.371.1813L,2006MNRAS.370..289B,2014MNRAS.439.1160R,IOT14,2015MNRAS.446.2380B,Latif_2016}.
In this scenario, the constraint on BH growth time is alleviated because the initial BH mass is larger,
but the duty cycle needs to be $O(1)$ \citep{Tanaka14}, which is difficult to achieve.
As an alternative, a massive BH could form via runaway stellar collisions in a globular cluster
\citep{1970ApJ...162..791S,2004Natur.428..724P,2006MNRAS.368..141F} and
in protogalaxies \citep{2008ApJ...686..801O,2009ApJ...694..302D,2015MNRAS.451.2352K,Yajima_2016}.
In any of these cases, seed BHs with $\sim 10^{2-5}~\msun$ still need to grow by subsequent accretion 
and mergers up to SMBHs with $\sim 10^9~\msun$ by $z> 6$ \citep[e.g.][]{TH09}.
The Soltan-Paczynski argument demands that most BH growth at $z\sim 2-3$ occurs in the accretion regime 
\citep{1982MNRAS.200..115S,2002MNRAS.335..965Y}.

\begin{figure}
\begin{center}
\includegraphics[width=75mm]{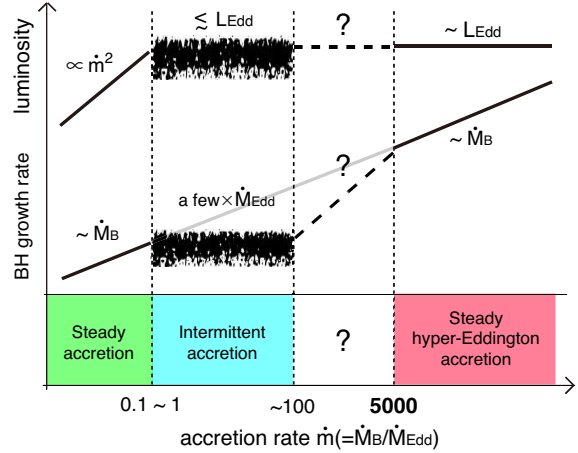}
\caption{A schematic overview of BH accretion solutions.
The characteristic behavior of the luminosity and BH growth rate are shown as a function of the Bondi accretion rate.
In the lowest $\mdot \la 0.1-1$ regime (green), the accretion flow results in an approximately adiabatic Bondi solution, 
where $L\propto \mdot^2$ and $\dot{M}\simeq \dot{M}_{\rm B}$.
In the intermediate $\mdot$ regime (blue), the accretion flow becomes unstable and highly time-dependent 
because of radiative feedback,
where $\langle L \rangle \la L_{\rm Edd}$ and $\langle \dot{M} \rangle \la {\rm a~few}\times \dot{M}_{\rm Edd}$ on time average.
In highest $\mdot$ regime (red), a new self-consistent solution of {\it steady hyper-Eddington accretion} is found 
in this paper.
In this new regime, the luminosity is limited to $\la L_{\rm Edd}$ because of photon trapping, 
and a very-high accretion rate is realized ($\dot{M}\simeq \dot{M}_{\rm B} \ga 5000~\dot{M}_{\rm Edd}$).
The region between the intermittent and hyper-Eddington accretion is still uncertain 
because the hydrodynamical instability (e.g. RT instability) might decrease the critical accretion rate.
}
\label{fig:sol_series}
\end{center}
\end{figure}

The basic physics of the accretion process was first studied for a spherically symmetric flow
without radiative feedback \citep{1952MNRAS.112..195B}. 
Solutions with radiative cooling and heating have been studied by many authors
\citep[e.g.][]{1973ApJ...180..531S,1990ApJ...354...64P,1990ApJ...354...83P,1991ApJ...383..250N,
2009ApJ...699...89C,2010ApJ...717..708C}.
With only two-body radiative processes included, the solutions are characterized by 
the normalized luminosity and accretion rate,
\begin{equation}
l\equiv \frac{L}{L_{\rm Edd}}~~~{\rm and }~~~
\mdot \equiv \frac{\dot{M}}{\dot{M}_{\rm Edd}}
\end{equation}
where $L_{\rm Edd}\equiv 4\pi cGM_{\rm BH}/\kappa_{\rm es}$ is the Eddington luminosity, 
$\kappa_{\rm es}$ is the electron scattering opacity,
and $\dot{M}_{\rm Edd}\equiv L_{\rm Edd}/c^2$ \citep{1985ApJ...288..428C}\footnote{
Note that this definition of $\dot{M}_{\rm Edd}$ is ten times smaller than that often used where 
the radiative efficiency is assumed to be $0.1$.}.
In Fig.~\ref{fig:sol_series}, we show schematically the characteristic behavior of the luminosity and BH growth rate 
as a function of the Bondi accretion rate, respectively.
For low $\mdot~(\ll 1)$, the accreting gas is optically thin to electron scattering,
and radiation processes (cooling, heating and pressure) are inefficient (green region in Fig.~\ref{fig:sol_series}).
Therefore, the accretion flow results in an approximately adiabatic Bondi solution,
where the accretion occurs within the Bondi radius $R_{\rm B}(\equiv GM_{\rm BH}/c_\infty^2)$, 
where $c_\infty$ is the sound speed of the ambient gas.
The luminosity from the accreting gas is as low as $l\la 10^{-7}$ 
\citep{1973ApJ...180..531S,1990ApJ...354...83P}.

On the other hand, for $\mdot \ga0.1$ (blue region in Fig.~\ref{fig:sol_series}), the hydrodynamical reaction to radiation processes 
has a major role in determining the properties of the accretion flow.
Heating by hot Compton radiation with $T_{\rm comp}\sim 10^8~\K$ from the gas near the BH 
strongly suppresses the accretion from the Bondi radius.
As a result, steady and self-consistent solutions exist only for $3\la \mdot \la 100$
and $l \la 10^{-2}$ \citep{1976ApJ...208L..61O,1990ApJ...354...64P,1990ApJ...354...83P,1991ApJ...383..250N}.
However, these solutions have been found to be unstable, resulting in a highly time-dependent flow
\citep[e.g.][]{1978ApJ...226.1041C,1996MNRAS.281.1183Z,2012MNRAS.427.2734N,2014ApJ...789..150G}.
We here examine a yet larger $\mdot$ regime and explore the existence of self-consistent solutions 
with {\it very} high $\mdot \gg 100$, where the Compton heating does not suppress the accretion,
which we hereafter refer to as the {\it hyper-Eddington} regime.

The accreting gas in fact forms a tiny disk around the central BH, 
even with small angular momentum.
Then, the gravitational energy is released via radiation through dissipative processes in the accretion disk
more efficiently than for spherically symmetric flows.
The typical efficiency is $\eta \sim 0.1$ for a thin disk around a non-spining BH \citep{1973A&A....24..337S}.
Several works have studied accretion flows by adding the radiation from a central disk by hand 
to a Bondi-like solution, and found that the accretion flow does not approach a steady state but becomes 
episodic due to ionization and heating
\citep{2001ApJ...551..131C, 2009ApJ...698..766M,2009ApJ...696L.146M,2011ApJ...737...26N,
2011ApJ...739....2P,2012ApJ...747....9P,2012MNRAS.427.2734N}.
In two-dimensional radiation hydrodynamical (RHD) simulations, 
these authors found that the time-averaged accretion rate onto a massive BH ($M_{\rm BH}\sim 10^{2-4}~\msun$)
is limited to the Eddington accretion rate ($\mdot \la \eta^{-1}$). 
Moreover, gas accretion from cosmological scales is strongly reduced by radiative heating,
because of the shallow gravitational potential of dark matter (DM) halos hosting the BHs \citep{2009ApJ...701L.133A}.
Again, {\it radiative heating can shut down gas supply from large scales and potentially quench the BH growth}
in this intermediate accretion rate domain (cf. blue region in Fig.~\ref{fig:sol_series}).

In addition, the radiation force can affect the properties of 
accreting gas near the BH.
For $\mdot >1$, the accretion luminosity released near the Schwarzschild radius
($R_{\rm Sch}\equiv 2GM_{\rm BH}/c^2$) is estimated as 
$\sim G\dot{M}M_{\rm BH}/R_{\rm Sch}~(\simeq \mdot L_{\rm Edd})$,
which naively exceeds the Eddington luminosity.
In fact, however, the luminosity in a spherical inflow does not exceed $L_{\rm Edd}$
because the photons are trapped and advected inward with the falling gas
within a characteristic ``trapping" radius before escaping via radiative diffusion 
\citep{1978MNRAS.184...53B}.
Under steady-state conditions, the trapping radius is given by
\begin{equation}
R_{\rm tr}\equiv \frac{\kappa_{\rm es}\dot{M}}{4\pi c}=\frac{\mdot}{2}R_{\rm Sch}.
\end{equation}
For high $\mdot~(\geq 2)$, the luminosity released above $R_{\rm tr}$
($\sim G\dot{M}M_{\rm BH}/R_{\rm tr}$) should thus be smaller than $L_{\rm Edd}$ \citep{1979MNRAS.187..237B}.
In the non-spherical case, RHD simulations (some of which include magnetic fields) suggest that
(1) the gas can accrete with $\mdot >1$ through a geometrically thick disk, 
where photon trapping occurs, and 
(2) the radiation flux is highly anisotropic and the luminosity can at least moderately
exceed the Eddington luminosity in the polar direction
\citep[e.g.][]{2005ApJ...628..368O, 2009PASJ...61L...7O, 2014ApJ...780...79Y, 2014ApJ...796...22F,
2014MNRAS.441.3177M,2015MNRAS.447...49S,2015PASJ...67...60T,2015MNRAS.454L...6M}.
A recent three-dimensional simulation \citep{2014ApJ...796..106J} found that 
high-$\mdot$ accretion is possible though magnetic-field advection of radiation in the disk 
makes photon trapping less efficient.
However, all of the above multi-dimensional RHD simulations have been limited 
to the region near the central BH ($\sim 100~R_{\rm Sch}\ll R_{\rm B}$).

In this paper, we investigate another domain and consider solutions with very high accretion rates ($\mdot \gg 100$),
examining self-consistent solutions of hyper-Eddington accretion
from large scales ($\ga R_{\rm B}$) onto massive BHs ($10^2\leq M_{\rm BH} \leq 10^6~\msun$).
We perform one-dimensional hydrodynamical simulations 
which include multi-frequency radiation transfer and non-equilibrium chemistry.
We first run simulations of the outer region which resolves the Bondi radius
($10^{-3}~R_{\rm B} \la r \la 10~R_{\rm B}$)
and find that hyper-Eddington accretion from the Bondi radius is realized without the negative effects of 
radiative feedback when $\mdot \ga 5000$ (red region in Fig.~\ref{fig:sol_series}).
Consistent with other authors, we find that
for $\mdot \la 5000$, gas accretion leads to episodic inflow due to radiative feedback and 
the average accretion rate is limited to $\langle \mdot \rangle \la 10$.
We study several cases for different parameters (e.g. $M_{\rm BH}$ and $n_\infty$) 
and clarify the physics of the transition from the episodic state to a steady state.
We find that this transition is well-explained by the comparison of the Bondi radius and the size of the ionized region 
(H$_{\rm II}$ region).
Second, for the hyper-Eddington solutions, 
further simulations of the inner region, which resolve the trapping radius 
($0.5~R_{\rm tr}\la r \la 10^{-3}~R_{\rm B}$), are conducted.
We confirm that when the emergent luminosity is limited to $\la L_{\rm Edd}$ because of photon trapping, 
radiation from the inner region does not affect the gas dynamics in the outer region.
Moreover, we examine whether our solution would connect with a hyper-Eddington accretion disk solution 
smoothly at small radii ($\la10^2~R_{\rm Sch}$).

We briefly highlight the primary new results in our paper compared to previous works.
For the outer-region simulations, our treatment is based on that of 2D-RHD simulations
\citep{2009ApJ...696L.146M,2011ApJ...739....2P,2012ApJ...747....9P}, 
where the radiative efficiency is assumed to be constant ($ \eta \sim 0.1$)
and the average accretion rate is $\langle \mdot \rangle \la \eta^{-1}$.
We update the model of the radiative efficiency including the photon trapping effect for $\mdot \gg 1$,
and find a new pathway to form the accretion solutions with a hyper-Eddington rate.
Recently, \cite{2015MNRAS.452.1922P} also studied solutions with high accretion-rates
with an 1D-RHD simulation using a 
gray-approximation around the Bondi radius, and noted the possibility of super-Eddington accretion.
Our new points are that combination of photon trapping and radiative recombination induces the transition to a steady state regime for
$\mdot \ga 5000$ and that the final steady state of the accreting gas is essentially an isothermal Bondi solution.
Moreover, we study the gas properties in the inner region, which is not resolved by \cite{2015MNRAS.452.1922P},
by solving the radiation transfer equations which include the photon trapping effect self-consistently.
With this addition, we show that our solutions may plausibly be connected to hyper-Eddington disk solutions
at the central region near the BH.

The rest of this paper is organized as follows. 
In \S2, we describe the methodology of our RHD simulation.
In \S3, we show the simulation results and discuss the necessary conditions required for
a steady, hyper-Eddington accretion flow to exist.
In \S4, we present a discussion of the growth of seed BHs in the early Universe, and
two other possible observational consequences and tests of our new solutions.
In \S5 and \S6, we discuss caveats of our simulations and summarize the main conclusions of this paper.


\section{Simulation method}

\subsection{Our strategy}
We here use the hydrodynamical simulation code (ZEUS, \citealt{1992ApJS...80..753S}) including 
multi-frequency radiation transfer, photoionization and heating,
and primordial chemical network.
We consider the simplest case, spherically symmetric accretion
\citep[e.g.][]{2011ApJ...739....2P}.
As a reference of the accretion rate, we define the Bondi rate for isothermal gas,
i.e. the specific heat ratio $\gamma=1$,
\begin{align}
\dot{M}_{\rm B}&=\pi e^{3/2}\rho_\infty R_{\rm B}^2 c_{\infty},\nonumber\\
&=\pi e^{3/2}\rho_\infty \frac{G^2M_{\rm BH}^2}{c_\infty^3},
\end{align}
where $\rho_{\infty}$ is the ambient gas density.
The sound speed is given by $c_\infty = \sqrt{k_{\rm B}T_\infty/(\mu m_{\rm p})}$, where
$T_\infty$ is the gas temperature, and $\mu$ is the mean molecular weight.
Throughout this paper, the reference Bondi radius and Bondi accretion rate are estimated 
by using $T_\infty=10^4$ K, $\gamma=1$ and $\mu=1.22$. 
But both of $\gamma$ and $\mu$ are solved for self-consistently in our simulations.

\begin{figure}
\begin{center}
\includegraphics[width=73mm]{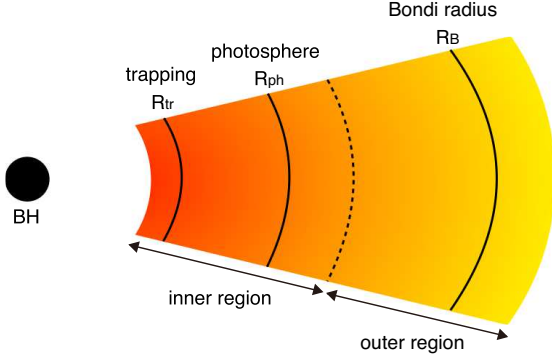}
\caption{A schematic picture of a spherically symmetric accretion flow onto a massive BH
at a hyper-Eddington accretion rate ($\mdot\gg 1$).
There are three characteristic scales: the Bondi radius ($R_{\rm B}$), photosphere ($R_{\rm ph}$),
and the trapping radius ($R_{\rm tr}$).
The dashed curve marks the boundary between the two regions simulated separately:
the outer region ($10^{-3}~R_{\rm B}\la r \la 10~R_{\rm B}$) and 
the inner region ($0.5~R_{\rm tr}\la r \la R_{\rm ph} \la 10^{-3}~R_{\rm B}$).
}
\label{fig:schematic}
\end{center}
\end{figure}

We need to consider gas dynamics over a wide range of spacial scales.
Fig. \ref{fig:schematic} shows a schematic picture of the flow.
In our case, the typical values of the Bondi and the trapping radius are 
\begin{equation}
R_{\rm B}\simeq 1.96\times 10^{18}~M_{\rm BH,4}T_{\infty,4}^{-1}~~{\rm cm},
\end{equation}
\begin{equation}
R_{\rm tr}\simeq 1.48\times 10^{12}~M_{\rm BH,4}\mdot_3~~{\rm cm},
\end{equation}
where $M_{\rm BH,4}\equiv M_{\rm BH}/(10^4~\msun)$, $T_{\infty,4}\equiv T_{\infty}/(10^4~\K)$
and $\mdot_3 \equiv \mdot/10^3$.
The required dynamical range of our simulation box is at least $6-7$ orders of magnitude in radius,
which is computationally prohibited.
Thus, we separate our simulation region into two regions of
an outer region ($10^{-3}~R_{\rm B}\la r \la 10~R_{\rm B}$) and 
an inner region ($0.5~R_{\rm tr}\la r \la 10^{-3}~R_{\rm B}$),
and first perform simulations of the outer region.
For the cases with $\mdot \gg 1$, we run several simulations 
by setting the gas properties (i.e. the density, thermal energy density, and velocity) at the inner-boundary 
of the outer-region simulation to the outer-boundary conditions of the inner-region simulation,
following each domain appropriate time resolution.
Finally, we obtain self-consistent solutions of the accretion flow onto a BH with 
hyper-Eddington accretion rates ($\mdot \ga 5000$)
by combining the results of the inner and outer regions.
Note that since solutions with $\mdot <5000$ do not approach a steady state 
due to radiative feedback in the outer region, 
we neither conduct the inner-region simulations nor obtain a fully self-consistent solution
in this unstable regime.

\subsection{Basic equations}
\label{sec:basic_eq}

The basic equations of hydrodynamics we solve are the following:
the equation of continuity
\begin{equation}
\frac{\partial \rho}{\partial t} + \frac{1}{r^2}\frac{\partial}{\partial r}{(r^2\rho v)}=0,
\end{equation}
the equation of motion
\begin{equation}
\rho \left(\frac{\partial {v}}{\partial t} + {v}\frac{\partial v}{\partial r}\right)
= -\frac{\partial p}{\partial r} -\rho \frac{\partial \Phi}{\partial r} + f_{\rm rad},
\end{equation}
where $\rho$ is the gas density, 
$v$ is the radial velocity (inflow; $v<0$), 
$p$ is the gas pressure, 
the gravitational potential with a general relativistic correction is set to
$\Phi= -GM_{\rm BH}/(r-R_{\rm Sch})$ \citep{1980A&A....88...23P},
and $f_{\rm rad}$ is the outward net radiation force in the radial direction.

We solve the energy equation including radiative cooling and heating,
\begin{equation}
\rho \left(\frac{\partial {e}}{\partial t} + {v}\frac{\partial e}{\partial r}\right)
= -p\frac{1}{r^2}\frac{\partial }{\partial r}(r^2v)-\Lambda+\Gamma,
\label{eq:energy_eq}
\end{equation}
where $e$ is the specific energy (erg g$^{-1}$).
The equation of state of the ideal gas is assumed as $p=(\gamma-1)\rho e$, 
where $\gamma=5/3$.
The first term of the right-hand side is the compressional heating term.
The last two terms are radiative cooling and heating, whose rates are 
$\Lambda$ and $\Gamma$ in units of erg s$^{-1}$ cm$^{-3}$.
The cooling rate is estimated as 
\begin{equation}
\Lambda = \Lambda_{\rm H}+\Lambda_{\rm He}+\Lambda_{\rm He^+}+
\Lambda_{\rm ff}+\Lambda_{\rm chem},
\end{equation}
where each term corresponds to the cooling rate associated with H, He, He$^+$ atoms,
free-free emission, and chemical reactions (see below).
For the outer-region simulation, we assume optically thin cooling rates of 
H atoms (Ly$\alpha$, $\Lambda_{\rm H}=\Lambda_{\rm Ly\alpha}^{\rm thin}$), 
He atoms ($1^1S$ state) and He$^+$ ions, and free-free transitions 
\citep{2007ApJ...666....1G}.
In the inner region, since the gas is opaque to Ly$\alpha$ photons, 
we solve the level population of H atoms ($2S$ and $2P$ state) including the Ly$\alpha $ trapping effect 
and estimate the cooling rate of two-photon emission \citep{O01}.
In addition to the H transitions, free-bound emission of H$^-$ 
(${\rm H}+{\rm e}^- \rightarrow {\rm H}^- + \gamma$) 
contributes as a cooling process. 
Thus, for the inner-region simulation, 
$\Lambda_{\rm H}=\Lambda_{\rm Ly\alpha}+\Lambda_{\rm 2ph}+\Lambda_{\rm H^-}$.
We show the details of our treatment of the Ly$\alpha$ trapping, continuum radiation cooling,
and opacity in the Appendix A.

We estimate the cooling rate by solving a chemical reaction network of metal-free gas,
which is composed of seven species (H, H$^+$, e$^-$, H$^-$, He, He$^+$, and He$^{++}$).
Since the reactions relevant to H$^-$ occur faster than the gas dynamical timescale,
the H$^-$ fraction is assumed to be in equilibrium (see Appendix A3).
The chemical reactions include photoionization, collisional ionization, radiative recombination
and collisional recombination of H, He and He$^+$.
Instead of considering photoionization by diffuse photons, we adopt the on-the-spot approximation
where the case A radiative recombination rate coefficient is replaced by that for case B.
To provide a stable, positive definite and first-order accurate solution of the chemical network, 
we use a method based on a semi-implicit formulation \citep{1997NewA....2..209A}.
The order of the updating is H, H$^+$, He, He$^+$, He$^{++}$ and e$^-$ \citep{2006ApJS..162..281W}.
For the inner-region simulation, inside the photosphere where all reactions are balanced,
the chemical abundances are determined by solving the Saha equations instead of 
the non-equilibrium reaction network.

To ensure the accuracy of solutions of the hydrodynamical equations coupled with radiative cooling/heating 
and primordial chemistry, the time step must be shorter than the Courant time (the Courant number is set to 0.1), 
cooling/heating time $t_{\rm cool}$ and chemical time $t_{\rm chem}$.
The cooling/heating time and chemical reaction time are given by
\begin{equation}
t_{\rm cool}=0.1~\frac{\rho e}{|\Lambda -\Gamma|},
\end{equation}
\begin{equation}
t_{\rm chem}=0.01~\frac{x_{\rm e}+0.001x_{\rm H}}{\dot{x}_{\rm e}},
\end{equation}
where $x_{\rm e}$ and $x_{\rm H}$ are the electron and neutral fraction
\citep{2006ApJS..162..281W,2008ApJ...672..287W}.
We set the time step to the lowest value among these timescales and 
integrate the energy equation (\ref{eq:energy_eq}) by an implicit method as well.

To solve the above basic equations, we employ spherical coordinates with a logarithmically-spaced grid in the radial direction:
the position of $i$-th grid is given by
$r_{i} = r_{\rm min} + \Delta r_0(\epsilon ^{i-1}-1)/(\epsilon -1)$ for $i=[1,N]$,
where $r_{\rm min}$ is the radius of the inner boundary, $\Delta r_0$ is the size of the inner-most grid-cell, 
$\epsilon~(= \Delta r_{i+1}/\Delta r_{\rm i})$ is the size ratio between consecutive grids,
$N$ is the number of grids, and 
the radius of the outer boundary $r_{\rm max}$ is given by $r_N$.
In our simulations, the size and layout of the coordinate grid is characterized by the four parameters 
($r_{\rm min}$, $r_{\rm max}$, $\epsilon$, and $N$).
The number of the grid cells is $N=700~(600)$ for the outer (inner) region simulations.
The value of $\epsilon$ is chosen to resolve the ionization front and 
the Bondi radius. Our fiducial value is $\epsilon = 1.01$.
We check the convergence of our results, changing $N$ from $600$ to $1800$ and $1.003\leq \epsilon \leq 1.01$.

\subsection{Radiation transfer}
\label{sec:RT}

To estimate the radiative heating rate $\Gamma$, ionization rate $\zeta$, and radiation pressure force,
we need to solve for radiation quantities in the accreting gas.
We here assume a steady state radiation field because the light crossing time ($\sim \tau r/c$, where $\tau$ is the optical depth) 
is much shorter than the time step of hydrodynamics
outside the trapping radius.
Thus, we have set the position of the innermost grid to $0.5~R_{\rm tr}$ in the inner-region simulation.  
In what follows, we describe how to solve the radiation transfer (RT) equations in the two regions, respectively.

\subsubsection{Outer region}

For the outer-region simulation, we follow the same treatment as
\cite{2011ApJ...739....2P,2012ApJ...747....9P}.
The RT equation is simply given by
\begin{equation}
\frac{1}{r^2}\frac{d}{dr}(r^2F_\nu) = 4 \pi \eta_\nu - \rho \kappa_\nu cE_\nu,
\end{equation}
where $F_\nu$ is the radiative flux, $E_\nu$ the radiation energy density, 
$\eta_\nu$ the emissivity, and $\kappa_\nu$ the absorption opacity.
Since the gas is transparent to photons with $h\nu <13.6$ eV, 
we solve the RT equation for ionizing photons ($h\nu \geq 13.6$ eV).
Inside the ionized region, the gas is optically thin even to ionizing photons and
thus $F_\nu \approx cE_\nu$.
Therefore, the radiation transfer equation is reduced to a simpler form:
\begin{equation}
\frac{1}{r^2}\frac{d}{dr}(r^2F_\nu) = 4 \pi \eta_\nu - \rho \kappa_\nu F_\nu,
\label{eq:RT_out}
\end{equation}
where the frequency range is $h\nu_{\rm min}=13.6$ eV $\leq h\nu \leq h\nu_{\rm max}=5$ keV.

We consider ionization and radiative heating due to bound-free absorption of H, He and He$^+$.
The ionization and heating rates are estimated so that the number of photons emitted along any line of sight 
will always equal the number of photo-ionizations in that direction over any time interval 
\citep{2006ApJS..162..281W,2008ApJ...682...49W} as
\begin{equation}
\zeta_{i}(r)=\int _{\nu_i}^\infty \frac{4\pi \hat{J}_\nu }{h\nu} \sigma_{i}~d\nu,\\
\end{equation}
\begin{equation}
n_{i}\Gamma_{i}=n_{i}\int _{\nu_i}^\infty \frac{4\pi \hat{J}_\nu}{h\nu} \sigma_{i} E_{{\rm heat},i}~d\nu,
\end{equation}
where $\hat{J}_\nu$ is the mean intensity which conserves the number of ionizing photons,
$\sigma_{i}$ ($i=$ H, He, He$^+$) is the bound-free cross section and $\nu_{i}$ is the ionization threshold,
$E_{{\rm heat},i}$ is the photoelectron's energy ($E_{{\rm heat},i}=h\nu - h\nu_{i}$).
The acceleration due to the momentum transfer from radiation is
\begin{equation}
f_{\rm rad}=\frac{\rho x_{\rm e}}{c}\int_{\nu_{\rm min}}^{\nu_{\rm max}} \kappa_{\rm es}F_\nu d\nu
+ \frac{\Gamma}{c}.
\end{equation}

In the outer-region simulation, for simplicity, we set a radiation source with a single power-law spectrum 
\begin{equation}
L_\nu = L_0\left(\frac{\nu}{\nu_{\min}}\right)^{-\alpha}~~(\nu\geq \nu_{\rm min}),
\end{equation}
and $L_\nu=0$ ($h\nu< h\nu_{\rm min}=13.6$ eV) as the inner boundary condition.
This assumption somewhat overestimates the effect of radiative feedback because 
this spectrum would be harder than that expected from self-consistent solutions (see below).
We here neglect the secondary ionization and heating due to high-energy electrons with $\gg 13.6$ eV
produced by the primary ionization.
Since this assumption is not valid for the case with a power-law spectrum, we underestimate the effect of ionizing radiation. 
However, radiation from the inner region would have a thermal spectrum with $\approx 10^4~\K$,
whose peak energy is $\simeq 2.4$ eV,
in the hyper-Eddington cases we are interested in 
(see \S\ref{sec:inner}).
We here set the power-law index to $\alpha=1.5$.

The normalization of the luminosity $L_0$ at the inner boundary 
is estimated from the bolometric luminosity given by
$L\equiv \int_{\rm \nu_{\rm min}}^{\rm \nu_{\rm max}} L_0d\nu=\eta \dot{M}c^2$, where $\eta$ is the radiative efficiency.
We here consider two models for $\eta$: (1) standard model ($\eta=$ constant) and 
(2) trapping model.
The standard model corresponds to radiation from a thin disk around the BH \citep{1973A&A....24..337S}.
For comparison with previous studies \citep[e.g.][]{2011ApJ...739....2P,2012ApJ...747....9P}, 
we consider several cases of $0.01\leq \eta \leq 0.3$.
As the trapping model, we adopt a simple prescription for the efficiency given by
\begin{equation}
\eta=\frac{3}{10+3 \mdot},
\label{eq:eta_tr}
\end{equation}
where $\eta \approx 0.3$ for $\mdot \ll 1$ and $\eta \approx \mdot^{-1}$ for $\mdot \gg 1$.
Since the trapping effect becomes important for $\mdot >2$, we set a transition point around $\mdot \simeq 3$.
For high $\mdot$, the luminosity has a maximum value of $L_{\rm Edd}$ ($l=1$)
because of the photon trapping effect.
We note that the choice of the maximum luminosity is motivated by 
\cite{1979MNRAS.187..237B}, who argued that the emergent luminosity is 
limited to $\la 0.6~L_{\rm Edd}$ even in a hyper-Eddington accretion phase.
As we discuss in \S3, physical processes to provide our main conclusion are not radiation pressure 
but photo-ionization heating.
Therefore, this choice does not change our main results significantly.

\subsubsection{Inner region}
\label{sec:RT_in}

In the inner region, a photosphere forms because of the high density and opacity.
We define the position of the photosphere $R_{\rm ph}$ as 
\begin{equation}
\tau_{\rm eff}(R_{\rm ph})=\sqrt{3\tau_{\rm abs}(\tau_{\rm abs}+\tau_{\rm scat})}=1,
\end{equation}
where $\tau_{\rm abs}$ and $\tau_{\rm scat}$ are the optical depth due to absorption 
(H$^-$ free-bound and H free-free transition) and scattering (H Rayleigh and electron scattering) processes.

As the initial conditions, we set uniform profiles of the density, temperature, velocity and neutral H fraction ($x_{\rm H}\simeq 1$),
whose values are given by those at the inner-boundary of the outer-region simulation.
If we run simulations under these conditions, the density increases at the center 
before a sufficiently large H$_{\rm II}$ region forms. 
Then, photoionization and recombination are tightly balanced at the boundary of the small H$_{\rm II}$ region
embedded by dense neutral gas.
As a result, the chemical reaction time becomes much shorter than the gas dynamical time, 
and thus we need a long computational time to reach the final steady state.
To avoid this problem, we adopt a value of the equilibrium opacity (Eqs. \ref{eq:eq_op1} and \ref{eq:eq_op2}) 
instead of the non-equilibrium opacity at the early stage of the simulation 
(note that the chemical reaction network is solved outside the photosphere; 
see \S\ref{sec:basic_eq}).
Because the equilibrium opacity is higher than the non-equilibrium opacity,
a photosphere forms in the H$_{\rm II}$ region before the density increases.
Since the chemical abundance inside the photosphere is given by the Saha equations, 
this approach saves significant computation time.
After one dynamical time at the outer-boundary ($t\geq 10^8$ s), 
we use the non-equilibrium opacity self-consistently
and obtain the final steady state at $t=2\times 10^8$ s.
We show the details about the treatment of opacities in Appendix A3.

Outside the photosphere ($r > R_{\rm ph}$), we can use the same RT equation as 
in the outer region (Eq. \ref{eq:RT_out}) for ionizing photons with $h\nu \geq 13.6$ eV.
For simplicity, we assume that the gas is optically thin against low-energy photons with $h\nu < 13.6$ eV
except Ly$\alpha$ photons ($h\nu=10.2$ eV).
We here neglect radiative cooling via other Lyman series and Balmer series photons
since their total cooling rate is comparable or smaller than that of two-photon emission
\citep{O01,2010ApJ...712L..69S,2012ApJ...750...66J}.

Inside the photosphere, radiation pressure dominates over the gas pressure
and photon trapping is important.
To capture these effects, we consider the following steady RT equations including the first-order correction 
in velocity $O(v/c)$ \citep{1982JCoPh..46...97M,1999ApJ...521..432L,2014ApJ...796..106J}
\begin{align}
\frac{1}{r^2}\frac{d}{dr}(r^2F_\nu) &= 4 \pi \eta_\nu - \rho \kappa_\nu c E_\nu\nonumber \\
&+\frac{\rho(\kappa_\nu -\kappa_{\rm es})v}{c}[F_\nu-v(E_\nu+P_\nu)],
\label{eq:RT1}
\end{align}
\begin{align}
\frac{dP_\nu}{dr} + \frac{3P_\nu-E_\nu}{r}= & -\frac{\rho (\kappa_\nu+\kappa_{\rm es})}{c} F_\nu
+\frac{\rho(\kappa_\nu+\kappa_{\rm es})v}{c}P_\nu\nonumber\\
&+\left(\frac{4\pi \eta_\nu}{c}+\rho \kappa_{\rm es} E_\nu \right)\frac{v}{c},
\end{align}
where all radiation moments are measured in the rest frame for an observer at infinity.
The last two terms of the right-hand side in Eq. (\ref{eq:RT1}), 
$\rho(\kappa_{\rm es} -\kappa_{\nu})(E_\nu+P_\nu)v^2/c$, 
are the second-order corrections in $(v/c)$.
These terms were included by \cite{1999ApJ...521..432L} to ensure the correct thermal equilibrium state 
in moving fluids when scattering dominates the opacity.

The Lorentz transformation between radiation moments in the rest frame ($E_{\nu},~F_{\nu},~P_{\nu}$)
and those in the fluid frame ($E_{0,\nu},~F_{0,\nu},~P_{0,\nu}$) is given by
$E_\nu =E_{0,\nu} +2v F_{0,\nu} / c^2$,
$F_\nu =F_{0,\nu} +v(E_{0,\nu}+P_{0,\nu})$
and
$P_\nu =P_{0,\nu} +2v F_{0,\nu} / c^2$.
Inside the photosphere, the radiation is thermalized and isotropic.
Thus, $E_{\nu,0}\approx 3P_{\nu,0}\approx 4\pi B_\nu(T)/c$,
where $B_\nu(T)$ is the Planck function, and the radiative flux in the fluid frame is approximated 
in the diffusion limit as $F_{\nu,0}\approx c\nabla E_{\nu,0}/\rho (\kappa_\nu +\kappa_{\rm es})\approx cE_{\nu,0}/\tau$.
Thus, the differences in $E_{\nu}$ and $P_{\nu}$ between the rest and fluid frame
is negligible because $(E_\nu-E_{\nu,0})=(P_\nu-P_{\nu,0})$ $\approx vF_{\nu,0}/c^2\approx (v/c)^2E_{\nu,0}/\tau$.
However, the difference in the radiative flux is not negligible inside the photosphere 
($\tau \gg 1$) because $F_\nu-F_{\nu,0}=v(E_{\nu,0}+P_{\nu,0})\approx \tau (v/c) F_{\nu,0}$ \citep{1984oup..book.....M}.

In the optically thick regime, we can approximate $E_\nu =3P_{\nu}(\approx E_{\nu,0})$ even in the rest frame.
Thus, $E_\nu \approx E_{0,\nu}= 4\pi B_\nu(T)/c$.
Then, the RT equations in the rest frame are 
\begin{align}
\frac{1}{r^2}\frac{d}{dr}(r^2F_\nu) &= \frac{\rho(\kappa_\nu -\kappa_{\rm es})v}{c}\left(F_\nu-4vP_\nu \right),
\end{align}
\begin{align}
F_\nu = -\frac{4\pi}{3\rho (\kappa_\nu+\kappa_{\rm es})}\frac{dB_\nu(T)}{dr} + 4vP_\nu.
\label{eq:F_lorentz}
\end{align}
We note that Eq. (\ref{eq:F_lorentz}) is identical to the Lorentz transformation of the radiative flux.
The first and second term corresponds to the diffusion and advection term, respectively.
After the frequency integration, the ratio of these terms is
\begin{align}
\frac{\rm diffusion}{\rm advection}
&=-\frac{cr}{\kappa_{\rm es} \rho rv}\frac{d\ln T}{d\ln r},\nonumber\\
&\approx \frac{4\pi r}{\kappa_{\rm es} \dot{M}}~\left( =\frac{r}{R_{\rm tr}}\right),
\end{align}
which means that the radiative flux in the rest frame has a positive (negative) value
at $r>R_{\rm tr}$ ($r<R_{\rm tr}$).

In order to solve the RT equations both inside and outside the photosphere, 
we need boundary conditions.
The photospheric luminosity $L_{\rm ph}$ and the effective temperature $T_{\rm ph}$
are eigen-values to be determined by the requirement that integration inward from the photosphere 
reaches acceptable central conditions.
The photospheric luminosity is determined by physical conditions inside the photosphere
(e.g. disk formation and/or shocks).
\cite{1979MNRAS.187..237B} estimated the maximum value of the emergent luminosity 
from a spherical accretion flow with a high $\mdot~(\gg 1)$
as $\simeq 0.6~L_{\rm Edd}$, based on analytical arguments.
Referring to this value, we produced several simulations for $0.01\leq L_{\rm ph}/L_{\rm Edd}\leq 1$,
and studied whether the accretion flows approach a steady state under these boundary conditions.
Then, we determine the physically-correct value of $L_{\rm ph}$ such that our solution smoothly connects with
a small central accretion disk, which would form within $r_{\rm min}$ (see \S\ref{sec:inner}).
Once setting $L_{\rm ph}$ and $R_{\rm ph}$, the bolometric radiative flux is given by 
$F_{\rm ph}=L_{\rm ph}/(4\pi R_{\rm ph}^2)$.
At the photosphere, $F_{\rm ph}=\sigma_{\rm SB}T_{\rm ph}^4$, where $\sigma_{\rm SB}$ is 
the Stefan-Boltzmann constant.
Assuming that the radiation has a thermal spectrum with $T_{\rm ph}$, 
we obtain $F_{\nu}(r=R_{\rm ph})=\pi B_\nu (T_{\rm ph})$.
Using this value, we solve the RT equation inside and outside the photosphere.
Finally, for the gas at $r<R_{\rm ph}$, we solve the temperature profile using 
\begin{equation}
\frac{dT}{dr} = -\frac{3\rho \kappa_{\rm R}}{16\sigma_{\rm SB}T^3}F + \rho \kappa_{\rm R}T\frac{v}{c},
\end{equation}
which is given by the frequency integration of Eq. (\ref{eq:F_lorentz}), 
i.e. $F\equiv \int F_\nu d\nu$, where $\kappa_{\rm R}$ is the Rosseland mean opacity.


\section{Results}

\subsection{Ordinary stellar-mass BH ($100~\msun$)}
\label{sec:PopIII}

In this section, we discuss the case of an ordinary stellar-remnant BH with $M_{\rm BH}=100~\msun$.
This corresponds to an accreting BH which forms through the gravitational collapse 
of a massive star.
As initial conditions of the outer region, we adopt a neutral uniform gas with 
$n_{\infty}=10^5~\cc$, $T_\infty=10^4~\K$ and $v=0$
for comparison with previous works.
The corresponding Bondi and Eddington accretion rates are 
$\dot{M}_{\rm B}=1.6\times 10^{-5}\msunyr$ and $\dot{M}_{\rm Edd}=2.2\times 10^{-7}\msunyr$
($\dot{M}_{\rm B}/\dot{M}_{\rm Edd}=72$), respectively.

\begin{figure}
\begin{center}
\includegraphics[width=78mm]
{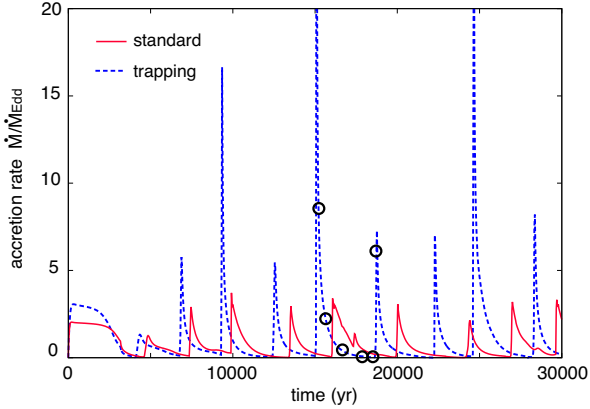}
\caption{
Time evolution of the gas accretion rate onto a massive stellar-remnant BH
($M_{\rm BH}=100~\msun$ and $n_\infty =10^5~\cc$; in the outer region).
Red solid and blue dashed curves correspond to the cases of the standard model with $\eta=0.3$
and the trapping model with $\eta=3/(10+3\mdot)$, respectively.
Open circles mark the six epochs at which we show radial profiles in Fig. \ref{fig:radial_eta}.
The corresponding Bondi accretion rate is $\dot{M}_{\rm B}\simeq 72~\dot{M}_{\rm Edd}$.
Even in the trapping model, where the luminosity does not exceed $L_{\rm Edd}$, 
the time averaged accretion rate is limited to $\sim \dot{M}_{\rm Edd}$.
This result demonstrates that gas accretion is quenched not by radiation pressure but by radiative heating.
}
\label{fig:t_Mdot}
\end{center}
\end{figure}

Fig.~\ref{fig:t_Mdot} shows the time evolution of the accretion rate (in Eddington units) for the standard model (red) with $\eta=0.3$, 
and the trapping model (blue) with $\eta=3/(10+3\mdot)$.
For both cases, the accretion is episodic.
During the brief burst phases, the rates are larger than the Eddington accretion rate,
whereas the long-term average rate is $\sim \dot{M}_{\rm Edd}$ for both cases.
The peak values for the trapping model are several times larger
because the maximum luminosity in the trapping model is limited to $\leq L_{\rm Edd}$
and thus radiation pressure is less efficient than in the standard model.
However, the average accretion rate is still $\sim \dot{M}_{\rm Edd}$ even in the trapping model.
This fact demonstrates that rapid accretion onto the BH is strongly quenched primarily not by radiation pressure 
but by radiative heating.

\begin{figure}
\begin{center}
\includegraphics[width=78mm]
{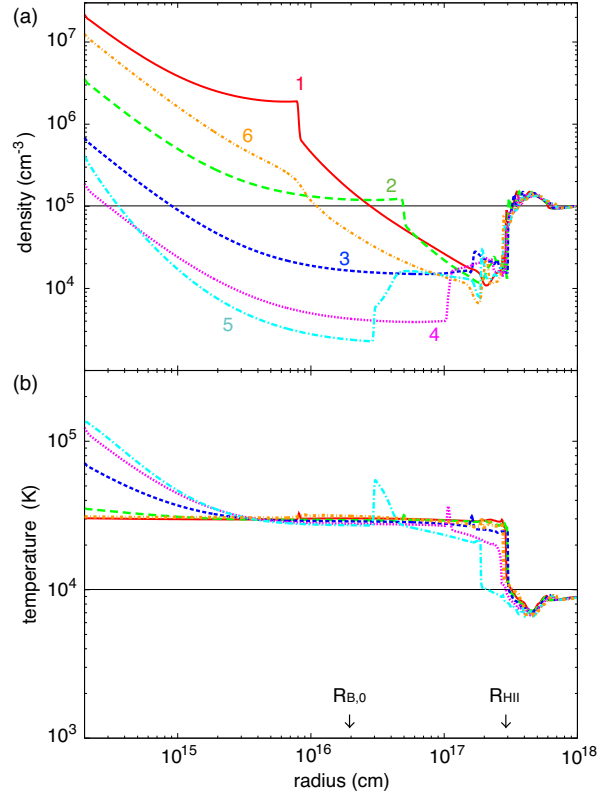}
\caption{Radial profiles of the (a) number density and (b) temperature of the accretion flow
over one cycle for $M_{\rm BH}=100~\msun$ and $n_\infty =10^5~\cc$ (in the outer region).
The curves show the profiles at (1) $t=1.52\times 10^4$ yr (red solid), (2) $1.57\times 10^4$ yr (green long-dashed),
(3) $1.66\times 10^4$ yr (blue short-dashed), (4) $1.77\times 10^4$ yr (magenta dotted),
(5) $1.84\times 10^4$ yr (light-blue long-dashed-dotted) and (6) $1.87\times 10^4$ yr (orange short-dashed-dotted)
The corresponding epochs are indicated by open circles in Fig. \ref{fig:t_Mdot}.
The horizontal lines are initial conditions ($n_\infty =10^5~\cc$ and $T_\infty=10^4~\K$) and 
the initial Bondi radius is $R_{\rm B,0}=2\times 10^{16}$ cm.
In this case where the size of the H$_{\rm II}$ region ($R_{\rm H_{II}}$) is larger than $R_{\rm B,0}$,
the gas dynamics is affected by radiation feedback (mainly heating), resulting in episodic accretion.
}
\label{fig:radial_eta}
\end{center}
\end{figure}

The physical reason why the accretion history becomes burst-like can be understood as follows.
Fig.~\ref{fig:radial_eta} shows radial profiles of the (a) number density and (b) temperature during one cycle
in the trapping model.
The corresponding epochs are indicated by open circles in Fig. \ref{fig:t_Mdot}.
Once the gas accretes onto the BH, the emergent radiation starts to ionize and heat the gas around the BH.
The hot H$_{\rm II}$ region expands until photoionization and recombination are balanced 
($R_{\rm H_{II}}\simeq 3\times 10^{17}$ cm). 
The H$_{\rm II}$ region becomes larger than the original Bondi radius, 
$R_{\rm B,0}=2\times 10^{16}$ cm (Fig. \ref{fig:radial_eta}b).
In the burst phase (curves 1--2 in Fig. \ref{fig:radial_eta}), 
the BH's gravity accelerates the inflow within a new sonic point in the H$_{\rm II}$ region 
($\simeq 3\times 10^{15}$ cm for $T\simeq 3\times 10^4~\K$), 
while the gas pressure (and partially the radiation pressure force) pushes the gas outward at $r\ga 10^{16}$ cm 
and the accretion rate is reduced.
As a result, in the quiescent phase (curves 3--5 in Fig. \ref{fig:radial_eta}), a density cavity forms within the H$_{\rm II}$ region, 
where the outward and inward gas pressure forces are balanced.
However, as the depletion of the hot gas proceeds and the outward gas pressure force decreases, 
a density bump forms inside the cavity at $3\times 10^{16}\la r\la 3\times 10^{17}$ cm 
(see the sharp feature in the curve 5 in Fig. \ref{fig:radial_eta}a).
Then, the gas is accelerated by the inward gas pressure (i.e. $dp/dr>0$) due to this bump.
Finally, the density bump falls into the central BH and produces burst-like accretion again
(curve 6 in Fig. \ref{fig:radial_eta}a).

A similar episodic behavior of the accretion rate for a constant radiative efficiency has been already described 
in previous works \citep[e.g.][]{2001ApJ...551..131C,2009ApJ...698..766M,2009ApJ...696L.146M,2011ApJ...739....2P,2012ApJ...747....9P}.
Although the radiation pressure force just after the burst-like accretion is stronger for the constant efficiency, 
the episodic behavior is due to radiative heating and leads to a similar behavior in the trapping model.

\subsection{Higher BH mass cases}
\label{sec:high_mass}

\begin{figure}
\begin{center}
\includegraphics[width=80mm]
{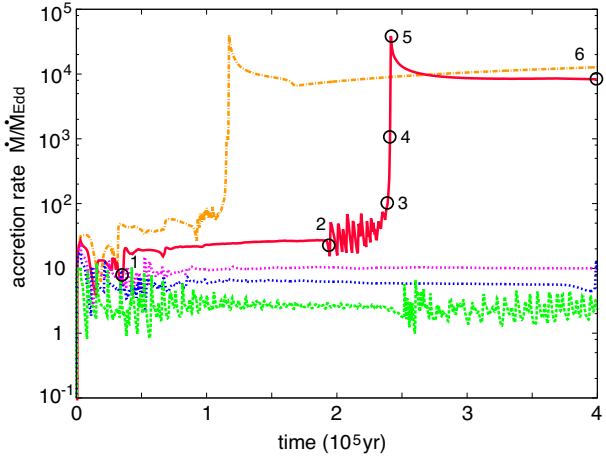}
\caption{Accretion rate history for a higher-mass BH (in the outer region). 
The curves show the cases for $M_{\rm BH}=10^3$ (green long dashed), $3\times 10^3$ (blue short dashed),
$5\times 10^3$ (magenta dotted), $10^4$ (red solid), and $2\times 10^4~\msun$ (orange dot-dashed).
The radiative efficiency is assumed to be $\eta=3/(10+3\mdot)$ (the trapping model).
The density of the ambient gas is $n_\infty =10^5~\cc$. 
Open circles mark the six epochs at which we show the radial profiles in Fig. \ref{fig:radial_high}.
For lower BH mass ($M_{\rm BH}<10^4~\msun$), the average accretion rate is limited to $\mdot \la 10$,
which is similar to that shown in Fig.~\ref{fig:t_Mdot}.
For higher BH mass ($M_{\rm BH}\geq 10^4~\msun$), a big jump of the accretion rate occurs and 
the accretion rate approaches a constant value, where $\dot{M}\simeq 8000~\dot{M}_{\rm Edd}$
($\dot{M}_{\rm B}\simeq 7200~\dot{M}_{\rm Edd}$).
}
\label{fig:t_Mdot_high}
\end{center}
\end{figure}

Next, we move onto the cases of more massive BHs.
Fig.~\ref{fig:t_Mdot_high} presents the accretion history for different BH masses 
($10^3\leq M_{\rm BH}\leq 2\times 10^4~\msun$).
We adopt the radiative efficiency of the trapping model (Eq. \ref{eq:eta_tr}).
For the lowest BH mass, the accretion occurs episodically as in the case of ordinary stellar-remnant massive BH as in \S\ref{sec:PopIII}.
With the BH mass increasing, the accretion rate becomes less time-dependent and the average accretion rate is 
$\sim 10~L_{\rm Edd}/c^2$, which corresponds to the case of a $10\%$ radiative efficiency many papers assume motivated by a thin disk
\citep[e.g.][]{1973A&A....24..337S}.
For the larger BH mass, we find a big jump of the accretion rate from $\mdot \sim 30$ to $\mdot \ga 10^4$
at $t\simeq 2.4\times 10^5~(1.15\times 10^5)$ yr for $M_{\rm BH}=10^4~\msun~(2\times 10^4~\msun)$.
After the transition for each case, the accretion rate approaches a constant value, 
which is identical to the Bondi accretion rate.

\begin{figure}
\begin{center}
\includegraphics[width=75mm]
{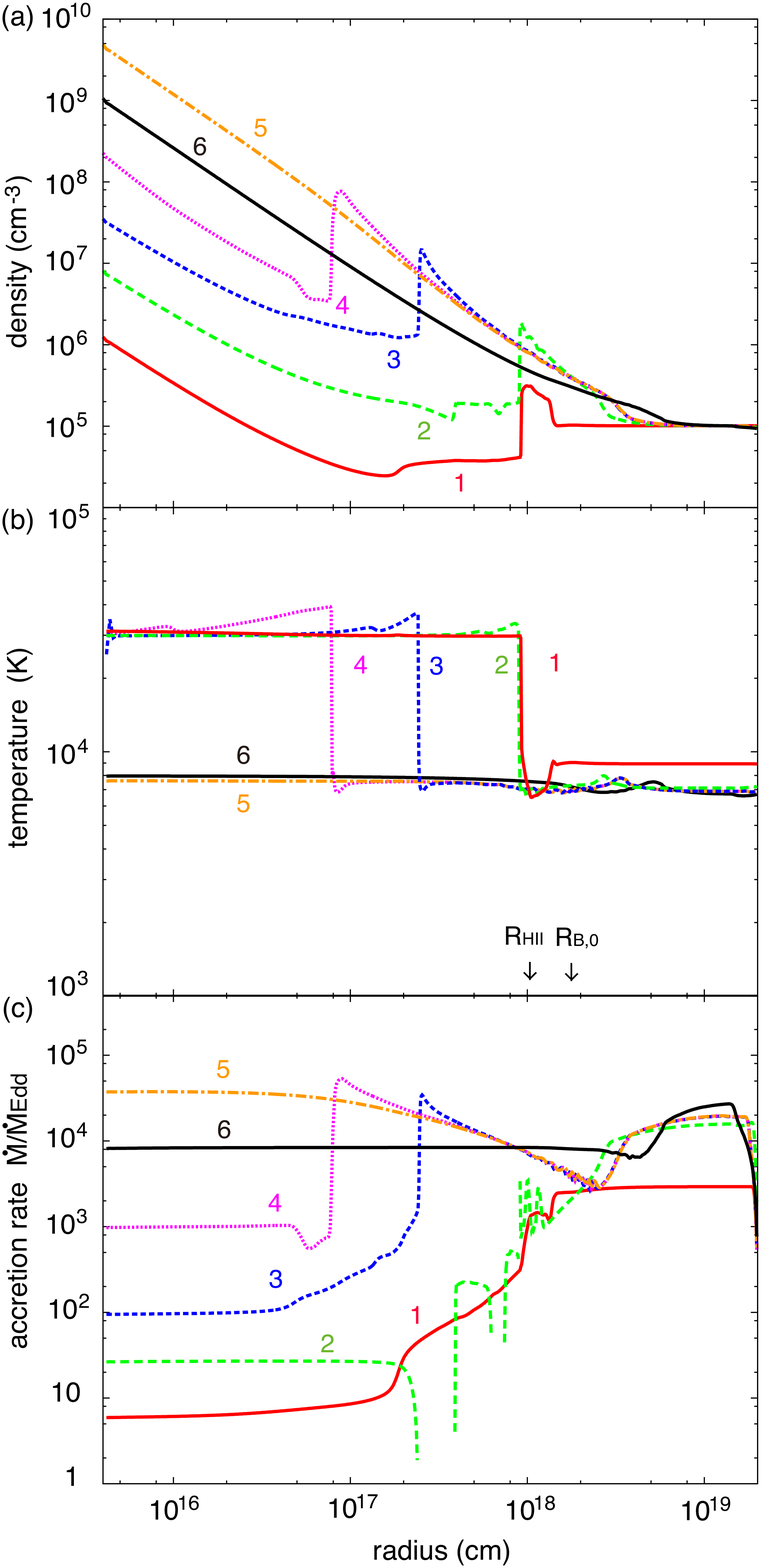}
\caption{Time evolution of radial profiles of the (a) number density, (b) temperature and
(c) inflow rate ($=-4\pi \rho vr^2$) for $M_{\rm BH}=10^4~\msun$ and $n_\infty=10^5~\cc$ (in the outer region).
The curves show the profiles at (1) $t=3.5\times 10^4$ yr (red solid), (2) $1.9 \times 10^5$ yr (green long-dashed), 
(3) $2.4\times 10^5$ yr (blue short-dashed), (4) $2.41\times10^5$ yr (magenta dotted), 
(5) $2.42\times10^4$ yr (orange dashed-dotted), and (6) $4 \times 10^5$ yr (black solid).
The initial Bondi radius is $R_{\rm B,0} = 2\times 10^{18}$ cm.
Note that in panel (c) for clarify, we do not show positive values (phase 2; green) at $r \ga 10^{17}$ cm.
In this case where the H$_{\rm II}$ region is confined within the Bondi radius ($R_{\rm H_{II}}<R_{\rm B,0}$), 
the big jump of the accretion rate occurs.
Finally, the accretion flow becomes a steady and isothermal Bondi solution with $\mdot \simeq 8000$.
}
\label{fig:radial_high}
\end{center}
\end{figure}

In Fig.~\ref{fig:radial_high}, we present radial profiles of the density, temperature, and inflow rate 
at different times for the case with the jump of the accretion rate 
($M_{\rm BH}=10^4~\msun$ and $n_\infty=10^5~\cc$).
In the early phase, the accretion history has several quiescent and burst phases as in 
the lower BH mass case (see \S\ref{sec:PopIII}).
In this case, however, the dense shell which developed at the edge of the H$_{\rm II}$ region pushes 
the hot ionized gas inward at $t\ga 2\times 10^5$ yr.
This results in the big jump of the accretion rate (phase 3--5 in Figs. \ref{fig:t_Mdot_high} and \ref{fig:radial_high}).
Note that the inward acceleration by the positive pressure gradient force is subdominant compared to 
that by the BH's gravity, in contrast to the burst-like accretion with a low accretion rate (see \S\ref{sec:PopIII}).
Fig. \ref{fig:radial_high}(b) shows that the size of the H$_{\rm II}$ region shrinks and it disappears.
Finally, the temperature profile is nearly isothermal with $T\simeq 8000~\K$.
The final profile of the inflow rate is almost constant (Fig. \ref{fig:radial_high}c), 
which is a steady and approximately isothermal Bondi solution with $\mdot\simeq 8000$.

We note the final fates with $10^3< M_{\rm BH} <10^4~\msun$,
where the accretion rates are almost constant at the end of the simulations.
For these marginal cases between the burst-like accretion ($M_{\rm BH}\leq 10^3~\msun$) and 
the hyper-Eddington accretion ($M_{\rm BH}\geq 10^4~\msun$),
a dense shell develops just outside the H$_{\rm II}$ region 
but radiative heating still suppresses the strong inflow.
We do not find a transition to a steady hyper-Eddington phase at least within the simulation time, 
which is $10-20$ times longer than the dynamical time at $R_{\rm B,0}$.
However, the dense shell might fall into the BH at some point in a longer-term simulation 
if the gravity slightly exceeds the outward force of gas pressure.
If in the case of density inversion flows, there is a non-linear density fluctuation due to 
the Rayleigh-Taylor instability (see \S\ref{sec:multi}), then accretion might increase
even for the marginal cases of $10^3< M_{\rm BH} <10^4~\msun$.

\begin{figure}
\begin{center}
\includegraphics[width=78mm]
{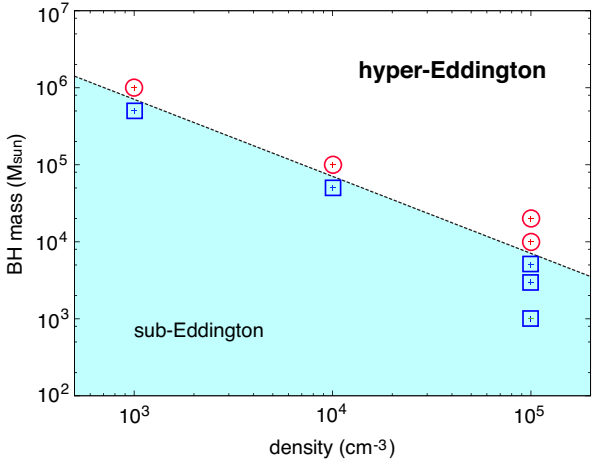}
\caption{Summary of the results for different values of the BH mass $M_{\rm BH}$ and 
number density of the ambient gas $n_\infty$.
Each symbol indicates whether the final result is hyper-Eddington accretion (circle), or
constant or episodic accretion at the rate of $\la (1-10)~\dot{M}_{\rm Edd}$ (square).
The dashed line marks the boundary between the two accretion modes: $M_{\rm BH,4}n_{\infty,5}=1/\sqrt{2}\simeq 0.71$.
The region above the boundary indicates the conditions required to realize steady hyper-Eddington accretion.
}
\label{fig:n_M_parameter}
\end{center}
\end{figure}

\subsection{Parameter dependence}

To discuss the conditions required to realize a steady hyper-Eddington accretion flow,
we conduct four additional simulations for different BH mass ($5\times 10^4 \leq M_{\rm BH}\leq 10^6~\msun$)
and number density of the ambient gas ($n_{\infty}=10^3$ and $10^4~\cc$).

The results for $n_\infty <10^5~\cc$ are similar to the cases shown in \S\ref{sec:high_mass}.
For a larger BH mass or higher density, the accretion rate jumps to the corresponding Bondi rate.
On the other hand, for smaller BH mass or lower density, the accretion rate is limited to $\la (1-10)~L_{\rm Edd}/c^2$.
Fig. \ref{fig:n_M_parameter} summarizes our results.
Each symbol indicates whether the final result is steady hyper-Eddington accretion (circle) or
sub-Eddington accretion ($\mdot \la {\rm a~few}$; square).
The dashed line marks the boundary between the two accretion modes: $M_{\rm BH,4}n_{\infty,5}=1/\sqrt{2}\simeq 0.71$.

\subsection{Analytic argument}
\label{sec:anal}

We here give a simple analytic argument for the conditions required for hyper-Eddington accretion.
As we explained in \S\ref{sec:high_mass}, the relation between the size of the H$_{\rm II}$ region and 
the Bondi radius is important to determine whether the transition to the hyper-Eddington accretion phase occurs.
For the lower BH mass case, the ionizing front propagates outside the Bondi radius and never shrinks 
($R_{\rm H_{II}}>R_{\rm B,0}$; see Fig. \ref{fig:radial_eta} b).
The radiation heating and pressure in this case can affect the gas dynamics at the Bondi radius.
Thus, the accretion is suppressed and $\mdot \sim 10$ (blue curve in Fig. \ref{fig:t_Mdot_high}).
For the higher BH mass, the H$_{\rm II}$ region is always confined within the Bondi radius 
($R_{\rm H_{II}}<R_{\rm B,0}$; Fig. \ref{fig:radial_high} b)
and shrinks dramatically at $t\ga 2.4\times 10^5$ yr.
As a result, the accretion from the Bondi radius cannot halt due to radiation feedback,
and the accretion flow becomes steady with $\mdot \simeq 8000$ 
(see Fig. \ref{fig:t_Mdot_high}).

The size of an H$_{\rm II}$ region in an uniform-density medium with $n_\infty$ is estimated as 
\begin{equation}
R_{\rm H_{II}}=\left(\frac{3Q_{\rm ion}}{4\pi \alpha_{\rm rec,B}n_\infty^2}\right)^{1/3},
\label{eq:RII}
\end{equation}
where $Q_{\rm ion}(\propto L)$ is the ionizing photon number flux (in units s$^{-1}$) 
and $\alpha_{\rm rec,B}$ is the H radiative recombination coefficient (case B).
For the trapping model ($L\leq L_{\rm Edd}$), the maximum value of $Q_{\rm ion}$ is
$L_{\rm Edd}/(h\langle \nu \rangle)\propto M_{\rm BH}$, where 
$h\langle \nu \rangle$ is the average energy of ionizing photons. We obtain
\begin{equation}
R_{\rm H_{II}, max}= 1.8\times 10^{18}~M_{\rm BH,4}^{1/3}~n_{\infty,5}^{-2/3}~T_{\rm H_{II},4}^{0.28}~~{\rm cm},
\end{equation}
where $T_{\rm H_{II},4}\equiv T_{\rm H_{II}}/(10^4~\K)$ is the temperature inside the H$_{\rm II}$ region 
and we set $h\langle \nu \rangle =13.6$ eV.
This value is larger by a factor of $\approx 2-3$ than the actual value 
because Eq. (\ref{eq:RII}) neglects the fact that the density profile has a steep slope 
($\rho \propto r^{-\beta}$; $0\la \beta \la3/2$) within $R_{\rm B}$.
We set $R_{\rm H_{II}}\approx R_{\rm H_{II},max}/2$.
Thus, the condition for the transition to the hyper-Eddington accretion 
($R_{\rm H_{II}}\la R_{\rm B}$) is written as
\begin{align}
M_{\rm BH,4}~n_{\infty,5}\ga 0.64~T_{\infty,4}^{3/2}~T_{\rm H_{II},4}^{0.42}
\end{align}
or 
\begin{align}
\mdot=\frac{\dot{M}_{\rm B}}{\dot{M}_{\rm Edd}}\ga 4.6\times 10^3~T_{\rm H_{II},4}^{0.42}.
\end{align}
For $T_{\rm H_{II},4}=3$, the critical accretion ratio is $\mdot \simeq 7.3\times 10^3$, 
which agrees well with the numerical simulation (see Figs. \ref{fig:t_Mdot_high} and \ref{fig:radial_high}).
In the following discussions (\S\ref{sec:apply} and \S\ref{sec:discussion}), 
we set the critical rate required for hyper-Eddington accretion to
$\dot{m}=5000$, which is given by the dashed line in Fig.~\ref{fig:n_M_parameter}.

\subsection{Inner-region simulations}
\label{sec:inner}

For the case of the hyper-Eddington accretion flow,
we further conduct simulations of the inner region, resolving the trapping radius.
Here, we study whether gas and radiation in the inner region 
affect or modify the hyper-Eddington solution of the outer region.
As we mentioned in \S\ref{sec:RT_in}, we have run several simulations for $0.01 \leq L_{\rm ph}/L_{\rm Edd} \leq 1$.
Since the choice of $L_{\rm ph}$ is still arbitrary, we attempt to determine the physically correct value of $L_{\rm ph}$ 
so that our solution smoothly connects with a small accretion disk well inside $R_{\rm tr}$.

\begin{figure}
\begin{center}
\includegraphics[width=73mm]
{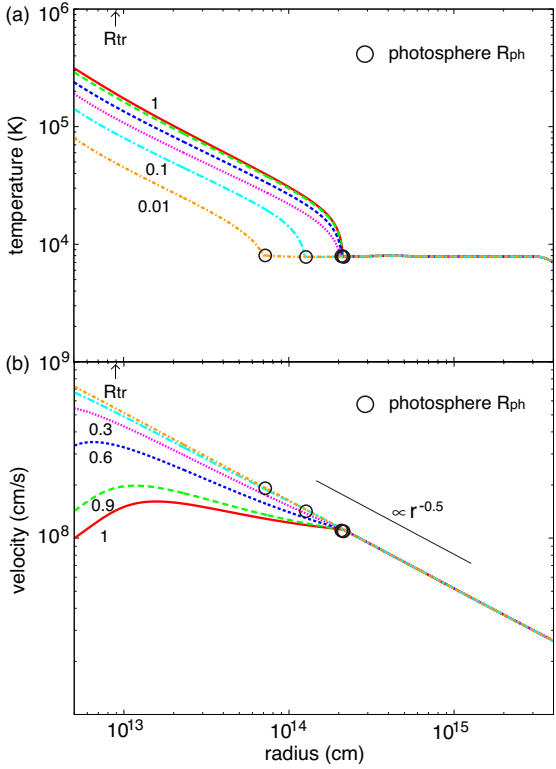}
\caption{The profile of the (a) gas temperature and (b) inflow velocity 
at the end of the simulation for $M_{\rm BH}=10^4~\msun$ 
and $n_\infty=10^5~\cc$ (in the inner region).
The curves correspond to the cases with the photospheric luminosity of 
$L_{\rm ph}/L_{\rm Edd}=1$ (red solid), $0.9$ (green long-dashed), $0.6$ (blue short-dashed), 
$0.3$ (magenta dotted), $0.1$ (light-blue long-dashed-dotted), and $0.01$ (orange short-dashed-dotted).
Open circles indicate the location of the photosphere $R_{\rm ph}$ for each case.
For $L_{\rm ph}\ga 0.3~L_{\rm Edd}$, the comoving luminosity exceeds the Eddington luminosity within $R_{\rm tr}$
(see Fig. \ref{fig:r_L_0.6}b), 
and the velocity begins to decelerate due to radiation pressure and would approach a settling solution ($v \propto r$).
A solution with $L_{\rm ph}\simeq (0.6 -1)~L_{\rm Edd}$ could connect with a hyper-Eddington accretion disk solution 
smoothly at small radii.}
\label{fig:r_L_v}
\end{center}
\end{figure}

First, we show the results of the inner-region simulation for the hyper-Eddington solution shown in Fig. \ref{fig:radial_high}
($M_{\rm BH}=10^4~\msun$ and $n_\infty =10^5~\cc$) with $L_{\rm ph}=0.6~L_{\rm Edd}$.
Fig. \ref{fig:r_L_v} presents radial profiles of the (a) temperature and (b) inflow velocity
at the end of the simulations ($t=2\times 10^8$ s) after a steady state is reached. 
For $L_{\rm ph}=0.6~L_{\rm Edd}$, 
the temperature profile is almost constant ($T\simeq 8000~\K$) outside the photosphere.
The size of the photosphere is $R_{\rm ph}=2.1\times 10^{14}$ cm and 
the effective temperature is $T_{\rm ph}=1.2\times 10^4~\K$ 
($\equiv [L_{\rm ph}/(4\pi \sigma_{\rm SB}R_{\rm ph}^2)]^{1/4}$).
Inside the photosphere, the gas temperature reaches a few $\times 10^5~\K$ at the inner boundary 
($r_{\rm min}=5\times 10^{12}$ cm).
The inflow velocity is close to free-fall ($\propto r^{-1/2}$) outside the photosphere, 
where $\rho \propto r^{-3/2}$,
while the radiation pressure force becomes important inside the photosphere,
the inflow velocity decreases and the density is slightly piled-up
within $R_{\rm tr}\sim 10^{13}$ cm.

\begin{figure}
\begin{center}
\includegraphics[width=72mm]
{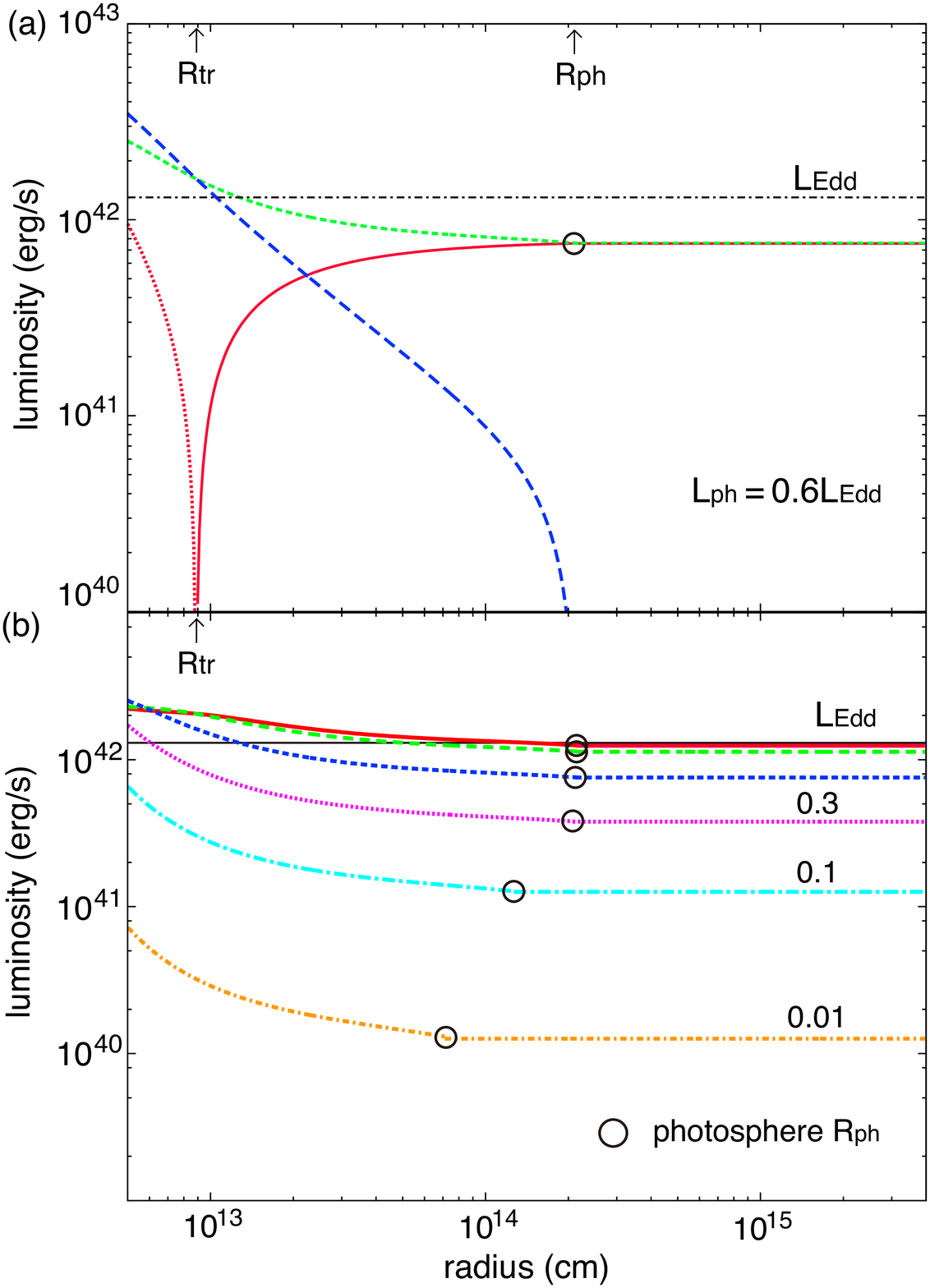}
\caption{The profile of the radiative luminosity at the end of the simulation
for $M_{\rm BH}=10^4~\msun$, $n_\infty=10^5~\cc$ (in the inner region).
In panel (a), where $L_{\rm ph}=0.6~L_{\rm Edd}$, red curve shows the luminosity in the rest (observed) frame; 
for $L_{\rm rest}\geq 0$ (solid) and $L_{\rm rest}<0$ (dotted).
Green short-dashed and blue long-dashed curve present the luminosity 
in the fluid's comoving frame and the advection luminosity, respectively.
The open circle indicates the position of the photosphere $R_{\rm ph}$.
The horizontal line is the Eddington luminosity.
The rest-frame luminosity (red curve) decreases inward from $R_{\rm ph}$ because of photon trapping, 
and becomes negative within the trapping radius $R_{\rm tr}$.
In other words, most of the energy released at $\la R_{\rm tr}$ is trapped within the inflow and falls into the BH,
while the leaked energy can reach outside the photosphere ($\simeq 0.6~L_{\rm Edd}$) by radiative diffusion (green curve).
Panel (b) shows the radial profile of the comoving luminosity with different values of $L_{\rm ph}$.}
\label{fig:r_L_0.6}
\end{center}
\end{figure}

Fig. \ref{fig:r_L_0.6} (a) shows the profile of the radiative luminosity at the final state ($t=2\times 10^8$ s)
for the case with $L_{\rm ph}=0.6~L_{\rm Edd}$.
Each curve corresponds to the luminosity $L_{\rm rest}=4\pi r^2 F$ in the rest (observed) frame (red),
$L_{\rm com}$ in the fluid (comoving) frame (green), and the advection luminosity $L_{\rm adv}$ (blue).
Note that $L_{\rm com}$ is given by the diffusion approximation, and $L_{\rm adv}=-16\pi r^2vP$
as shown in Eq. (\ref{eq:F_lorentz}).
If the gas were static, i.e. $v=0$, the comoving luminosity would be identical to the rest-frame luminosity.
In fact, however, the rest-frame luminosity decreases towards the center from the photosphere
because radiation is trapped in the accretion flow due to electron scattering.
Within the trapping radius, the advection luminosity exceeds the comoving luminosity, and thus
the rest-frame luminosity becomes negative (red dotted).
The comoving luminosity increases towards the center with the advection luminosity and
reaches the Eddington luminosity at $R_{\rm tr}$.
Thus, within $R_{\rm tr}$, the inflow is decelerated by the radiation force.

We can give a different interpretation concerning the behavior of the luminosity in Fig. \ref{fig:r_L_0.6} (a).
Suppose that the kinetic energy is converted to radiation energy at $R_{\rm diss}(\la r_{\rm min})$
by some dissipative processes (e.g. by viscous processes in a disk or by shocks).
Then, the maximum value of the luminosity (in the fluid frame) can be 
$GM_{\rm BH}\dot{M}/R_{\rm diss}=L_{\rm Edd}(R_{\rm tr}/R_{\rm diss})> L_{\rm Edd}$.
The rate of energy release exceeds the Eddington luminosity, but 
most of the energy is trapped within the inflow and falls into the BH.
The leaked energy can reach outside the photosphere only by radiative diffusion, 
and the photospheric luminosity results in $0.6~L_{\rm Edd}$.

Next, we consider cases with different values of the assumed photospheric luminosities 
($0.01 \leq L_{\rm ph}/L_{\rm Edd}\leq 1$).
We show the radial profile of the comoving luminosity (Fig. \ref{fig:r_L_0.6}b) at the end of the simulations
as well as those of the inflow velocity and temperature (Fig. \ref{fig:r_L_v}). 
For all cases, the values of luminosity increase towards the center.
For $L_{\rm ph}\ga 0.3~L_{\rm Edd}$, the comoving luminosity exceeds the Eddington luminosity 
within the trapping radius, and the velocity begins to decelerate due to radiation pressure
and would approach the so-called {\it settling solution} ($v\propto r$).
The temperature inside the photosphere is higher for larger $L_{\rm ph}$,
since a larger temperature gradient produces a higher luminosity (i.e. diffusion approximation).
These results are consistent with previous works, which discussed steady-state solutions for 
an optically-thick accretion flow 
\citep[e.g.][]{1980ApJ...240..235G,1982MNRAS.199..833F,1984ApJ...284..394V,
1986ApJ...308..755B,1991ApJ...376..234H}.
These studies have concluded that the velocity deceleration is inevitable 
unless the photospheric luminosity is close to a specific eigen-value,
which is estimated to be $\simeq 3.5\times 10^{-5}L_{\rm Edd}$ for $M_{\rm BH}=10^4~\msun$ and $\mdot= 8000$.
Thus, the accretion flows would be settling solutions near the center for all our cases 
($0.01\leq L_{\rm ph}/L_{\rm Edd}\leq 1)$, but the deceleration point for smaller $L_{\rm ph}$ shifts to a smaller radius,
which is not resolved in our simulations for $L_{\rm ph}\la 0.3~L_{\rm Edd}$.

The profiles of the hyper-Eddington accretion solution within $R_{\rm ph}$ depend on the choice of $L_{\rm ph}$,
which is still arbitrary and is determined by matching inner boundary conditions.
We here clarify the values of $L_{\rm ph}$ for which the accretion flow would smoothly 
connect with a compact thick disk, which is formed at the bottom of the settling solution.

In simulations of an accretion disk with $\mdot \simeq 9800$
\citep{2015MNRAS.447...49S}, the surface density was found to be
$\Sigma_{\rm disk}\simeq 2\times 10^{6}~{\rm g~cm}^{-2}$ 
at $50~R_{\rm Sch}$, within which the net mass flux is negative (i.e. inflow).
The mid-plane temperature is $T_{\rm disk}\simeq 2\times 10^7~\K$ for 
$1600\la \mdot \la 8000$.
In the most extreme case, our solution with $\mdot \simeq 8000$ and $L_{\rm ph}=L_{\rm Edd}$ 
approaches a settling solution ($v\propto r$; see Fig. \ref{fig:r_L_v}b).
The dynamical time ($t_{\rm dyn}=r/|v|\simeq 5\times 10^4$ s) at $r<r_{\rm min}$ becomes constant and
$|v|=3.0\times 10^6$ cm s$^{-1}$ at $50~R_{\rm Sch}$ for this solution.
The surface density is estimated as $\Sigma = \dot{M}h/(-2\pi vr^2) \sim 2.0\times 10^6$ g cm$^{-2}$, 
where we set the disk thickness to $h/r\simeq 0.5$.
In addition, since the radiation pressure dominates within the trapping radius ($T\propto \rho^{1/3}$) and 
the density follows $\rho \propto r^{-3}$ in the steady settling solution ($v\propto r$),
we also find $T\propto r^{-1}$.
Thus, we can extrapolate $T\simeq 1.0\times 10^7~\K$ at $50~R_{\rm Sch}$ 
from the simulation result in Fig. \ref{fig:r_L_v} (a).
We conclude that our solution for $L_{\rm ph}=L_{\rm Edd}$ plausibly links to the outer edge of the smaller accretion disk.

For smaller values of $L_{\rm ph}$ ($<0.3~L_{\rm Edd}$), the dynamical time at the central region would be shorter than $10^4$ s, 
because the velocity deceleration would occur further inside $r_{\rm min}$.
Then, we can estimate $\Sigma < 4.1\times 10^5$ g cm$^{-2}$ and $T < 6.4\times 10^6~\K$ at $50~R_{\rm Sch}$,
which are smaller than the values of the numerical simulations with similar $\mdot$.
We conclude that a high-luminosity solution with $L_{\rm ph}\simeq (0.6-1)~L_{\rm Edd}$ among our solutions 
can be physically correct\footnote{For moderately larger values of $L_{\rm ph}$ ($>L_{\rm Edd}$), 
the radiation pressure would dominate over the BH's gravity inside the photosphere.
Thus, the accretion flow could not reach the outer edge of the disk nor connect with the hyper-Eddington disk solution.
However, we note that the maximum value of the accretion luminosity released at 
$R_{\rm tr}$ is $L_{\rm Edd}(=GM_{\rm BH}\dot{M}/R_{\rm tr})$
unless extra dissipation due to e.g. magnetic fields occurs (see \S\ref{sec:discussion}).}.
The gas is accelerated in the disk to the speed of light even with a subsonic velocity at the outer edge of the disk
\citep[e.g.][]{1988ApJ...332..646A,2009ApJS..183..171S,2011A&A...527A..17S,2014ApJ...796..106J}.
\cite{2014ApJ...796..106J} recently presented a similar, related study for lower $\mdot(\simeq 200)$,
which is too low to match our hyper-Eddington solutions.


\section{Consequences of hyper--Eddington accretion}
\label{sec:apply}

We next discuss three possible applications of the hyper-Eddington accretion solution: 
(1) growth of massive seed BHs in the early Universe, and 
observational signatures of rapidly accreting massive BHs as (2) Ly$\alpha$ emitters without X-rays 
and (3) bright ultra-luminous infrared sources.

\subsection{Rapid growth of seeds of SMBHs}

As a first application, we consider rapid growth of BHs in the early Universe.
Many previous studies \citep[e.g.][references therein]{2012RPPh...75l4901V}
have discussed the possibility of super-Eddington accretion to explain the existence of 
SMBHs observed at $z\simeq 6$.
We here simply assume a spherically symmetric gas distribution around the BH\footnote{The previous works suppose 
that the BH is embedded in a galactic disk ($\gg 1$ pc) and/or a compact disk ($\ll 1$ pc).
However, the mechanism of angular momentum transfer of the gas from the galactic scale to the BH scale
is still debated \citep[e.g.][]{2010MNRAS.407.1529H}.
In fact, the conditions for hyper-Eddington accretion depend on assumptions about the angular momentum transfer
(see also \citealt{2014Sci...345.1330A} for a possible solution).}
and show an impact of the new effect from our simulations, i.e. 
{\it the hyper-Eddington accretion phase is possible only if $\mdot > 5000$ is achieved,
otherwise the accretion rate is limited to $\mdot \la 10$ by radiative heating.}
Again, we note that our definition of $\dot{M}_{\rm Edd}$ is ten times smaller than that the previous papers used.
This condition is much more severe than the previous works supposed ($\mdot \ga {\rm a~few}\times 10$).

We consider an accreting massive BH embedded in a dense gas cloud at the center of 
a DM halo with a virial temperature of $\ga 10^4~\K$.
According to cosmological simulations of high-$z$ protogalaxies, 
a gas density profile following $\propto r^{-2}$ develops due to efficient radiative cooling
\citep[e.g.][]{2007ApJ...665..899W}.
For specificity, we estimate the density at the outer boundary of the halo by 
the minimum-energy truncated isothermal sphere model
\citep{1999MNRAS.307..203S,2001MNRAS.325..468I,2007MNRAS.375..881A} as
\begin{align}
n_{\rm vir}\simeq 0.07~\cc 
\left(\frac{1+z}{21}\right)^3.
\end{align}
The virial radius and virial temperature are 
\begin{align}
R_{\rm vir}\simeq 7.9\times 10^2~M_{\rm h,8}^{1/3}~{\rm pc} 
\left(\frac{1+z}{21}\right)^{-1},
\end{align}
\begin{align}
T_{\rm vir}\simeq 1.9\times 10^4~M_{\rm h,8}^{2/3}~\K
\left(\frac{1+z}{21}\right),
\end{align}
respectively.
The gas density profile is given by
\begin{align}
n(r)&=f_n n_{\rm vir} \left(\frac{r}{R_{\rm vir}}\right)^{-2},\nonumber\\
&\simeq 10^3~T_{\rm vir,4}~\cc \left(\frac{r}{10~{\rm pc}}\right)^{-2}
\left(\frac{f_n}{4}\right),
\label{eq:density_cosmo}
\end{align}
where $f_n$ is a numerical factor. 
We choose $f_n=4$ so that the density profile agrees with
cosmological simulations of atomic cooling halos with $T_{\rm vir}\simeq 10^4~\K$
\citep{2008ApJ...682..745W, 2010MNRAS.402.1249S,2014MNRAS.439.1160R}.
Thus, the density at the Bondi radius is 
\begin{equation}
n (R_{\rm B})\simeq 9.5\times 10^5~M_{\rm BH,4}^{-2}T_{\infty,4}^2 T_{\rm vir,4}~\cc,
\label{eq:density_RB}
\end{equation}
and the ratio of the Bondi rate to the Eddington rate is
\begin{align}
\frac{\dot{M}_{\rm B}}{\dot{M}_{\rm Edd}}
=6.9\times 10^4~M_{\rm BH,4}^{-1}T_{\infty,4}^{1/2}T_{\rm vir,4}.
\label{eq:BH_acc_cosmo}
\end{align}
From the necessary conditions for the hyper-Eddington accretion ($\mdot>5000$), we obtain
\begin{equation}
M_{\rm BH}\la 1.4\times 10^5~T_{\infty,4}^{1/2}T_{\rm vir,4}~\msun.
\label{eq:BHmass_cosmo}
\end{equation}
In fact, the gas inflow from the virial radius supplies sufficient gas
to the Bondi radius at the rate of $\sim 0.1-1~\msunyr$ \citep[e.g.][]{2015MNRAS.452.1026L},
which is high enough to maintain the hyper-Eddington accretion phase 
($5000~\dot{M}_{\rm Edd}\simeq 0.11~M_{\rm BH,4}~\msunyr$).
Thus, the BH mass can reach at least $M_{\rm BH}\ga 10^5~\msun$ via hyper-Eddington accretion,
independent of the initial BH mass.

Note that this argument is based on the assumption that the gas density profile ($\propto r^{-2}$) 
inside the halo would not be broken by stellar feedback before the BH forms.
This may be valid for two scenarios of seed BH formation: collapse of 
(1) Pop III stars ($\sim 100~\msun$) and (2) supermassive stars ($\sim 10^5~\msun$).
In the Pop III case, although the stars emit strong ionizing radiation and produce energetic explosions, 
the feedback would not be strong enough to blow the gas in the halo away, 
if the halo is as massive as that with $T_{\rm vir}\ga 10^4~\K$
\citep[e.g.][]{2004ApJ...613..631K,2004ApJ...610...14W,2005ApJ...630..675K}.
In the supermassive-star case, stellar feedback would be also inefficient
because the star has a bloated cold envelope with $\sim 5000~\K$ and 
directly collapses to a BH without mass-loss and explosions
(\citealt{2012ApJ...756...93H,2013MNRAS.431.3036I,
2013ApJ...778..178H,2013A&A...558A..59S}, but see also \citealt{2014ApJ...790..162C}).
Therefore, for these two seed BH models, hyper-Eddington accretion occurs 
in an atomic cooling halo with $T_{\rm vir}\simeq 10^4~\K$, where the density profile follows $\propto r^{-2}$.
This of course requires that there would be essentially no metal pollution due to prior massive star formation
(e.g. due to negative feedback from a Lyman-Werner radiation background).

We briefly mention the subsequent BH growth via hyper-Eddington accretion 
in more massive DM halos with $T_{\rm vir} \gg 10^4~\K$.
Since the gas density is higher in larger halos, rapid BH growth can be triggered by major mergers of galaxies.
However, the hyper-Eddington accretion phase would be quenched by non-spherical effects for higher 
DM and/or BH mass in the following way \citep[e.g.][]{2005ApJ...633..624V,2012MNRAS.425.2892W,2015ApJ...804..148V}.
In a massive DM halo, the gas cloud would flatten or form a thin disk 
because the DM gravity is stronger than the gas pressure support (i.e. $T_{\rm vir} \gg T_\infty \sim 10^4~\K$).
Then, the scale height of the disk becomes smaller than the Bondi radius with the BH mass increasing.
Therefore, the quasi-spherical accretion is no longer a good approximation and our results would not be directly applicable.
To explore the nature of a non-spherical accretion in such systems is left for future investigations.

Once seed BHs grow up to $\ga 10^{5}~\msun$ by hyper-Eddington accretion, 
they do not need more rapid growth phases to form high-$z$ SMBHs with $\ga 10^9~\msun$.
Even if the rapid growth is prohibited in massive DM halos with $T_{\rm vir}\gg 10^4~\K$, 
the BHs can grow at a sub-Eddington accretion rate.
The characteristic growth timescale from $10^5~\msun$ to $10^{9}~\msun$ is estimated as 
$\simeq 0.46$ Gyr, where the radiative efficiency is set to a constant value of $\eta =0.1$.
The growth time is shorter than the cosmic time from $z=20$ to $z=7$ ($\simeq 0.57$ Gyr).
We could explain the existence of high-$z$ SMBHs by combining hyper-Eddington accretion at the early stage of $z\simeq 20$
and the subsequent moderate growth \citep[e.g.][]{2012ApJ...745L..29D,2013ASSL..396..293H,Tanaka14}. 
The Soltan-Paczynski argument \citep{1982MNRAS.200..115S,2002MNRAS.335..965Y} 
assures that most of the mass in current epoch BHs grew via radiatively efficient near-Eddington accretion phases
in normal active galactic nuclei (AGN).

\subsection{Observational signature as bright sources}

Here, we discuss possible observational signatures of BHs accreting at the hyper-Eddington regime.
We here consider two cases -- with and without dust grains.
Depending on the metallicity, the accretion flow would be a bright source of Ly$\alpha$ emission ($Z\simeq 0$)
or infrared emission ($Z\simeq 10^{-3}~\zsun$).
For both cases, the object could be dim in the X-ray bands because of efficient photon trapping.

\subsubsection{Ly$\alpha$ emitters without X-rays}

Ly$\alpha$ emission is a useful probe for studying the intergalactic medium at the epoch of cosmic reionization.
Among high-$z$ galaxies, Ly$\alpha$ emitters (LAE) have prominent Ly$\alpha$ emission,
typically $L_{\rm Ly\alpha}\sim 10^{42-44}$ erg s$^{-1}$ 
\citep{2003ApJ...582...60O,2004AJ....127..563H,2004ApJ...617L...5M,2006ApJ...648....7K,
2006PASJ...58..313S,2008ApJS..176..301O,2010ApJ...723..869O,2010ApJ...724.1524O,
2012ApJ...744...83O}.
Basically, most LAEs are considered to be low-mass, star-forming galaxies, which could be building blocks of 
the massive galaxies at lower-redshift epochs.
However, LAEs have a variety of properties \citep[e.g.][]{2014ApJ...786...59H} and 
some LAEs do not have a stellar continuum detection, and/or have rest-frame equivalent
width in excess of 240~\AA, which cannot be explained by a normal stellar population with a Salpeter IMF
\citep[e.g.][]{2002ApJ...565L..71M,2004ApJ...617..707D,2006PASJ...58..313S}.
The origin of the strong Ly$\alpha$ emission in these cases remains uncertain.
As a possible process producing strong Ly$\alpha$ emission, we propose rapid accretion onto a massive BH 
($M_{\rm BH}\sim 10^{5-6}~\msun$).

In the very low-metallicity case, the accretion flow onto the BH releases a fraction of 
the gravitational energy as Ly$\alpha$ emission.
Then, the temperature is kept at $\simeq 8000~\K$ outside the photosphere
by balancing the compression with Ly$\alpha$ cooling.
The total Ly$\alpha$ luminosity from the accretion flow is estimated as \citep{2000ApJ...537L...5H}
\begin{align}
L_{\rm Ly\alpha}&=\int_{R_{\rm Ly\alpha}}^{R_{\rm B}} 4\pi r^2 n({\rm H})n({\rm e}) \bar{\Lambda}_0 dr,\nonumber\\
&=\frac{\bar{\Lambda}_0\dot{M}^2x_{\rm e}}{8\pi m_{\rm p}^2GM_{\rm BH}}\ln \left(\frac{R_{\rm B}}{R_{\rm Ly\alpha}}\right),
\end{align}
where $\bar{\Lambda}_0=2\times 10^{-25}$ erg s$^{-1}$ cm$^3$ is the value of the Ly$\alpha$ cooling function 
at $T\simeq 10^4~\K$, $x_{\rm e}\simeq 10^{-4}$ is the electron fraction of the accretion flow outside the Ly$\alpha$ 
trapping radius $R_{\rm Ly\alpha}$ (see Appendix A1).
Using Eq. (\ref{eq:lya_tr}), we obtain 
\begin{equation}
L_{\rm Ly\alpha}\simeq 2\times 10^{38}~{\rm erg~s^{-1}} M_{\rm BH,4}^3n_{\infty,5}^2.
\label{eq:L_Lya}
\end{equation}
Thus, for massive accreting BHs with $\sim 10^{5-6}~\msun$ embedded in dense clouds ($n_\infty \sim 10^5~\cc$), 
the Ly$\alpha$ luminosity is as bright as $\simeq 10^{42-44}$ erg~s$^{-1}$,
which agrees with the typical luminosity of LAEs.
Ly$\alpha $ emission with a flux of $> 2\times 10^{-18}(L_{\rm Ly\alpha}/10^{43}~{\rm erg s^{-1}})$ erg s$^{-1}$ cm$^{-2}$ 
can be detected out to redshift $z\simeq 20$ by the Near-Infrared Spectrograph on the James Webb Space Telescope (JWST) 
with an exposure time of $10^5$ s in the $R=1000$ grating mode\footnote{http://www.stsci.edu/jwst/instruments/nirspec/sensitivity}.
Note that the argument would not be valid for $M_{\rm BH} \gg 10^6~\msun$
because the Ly$\alpha$ luminosity must be smaller than the Eddington luminosity
in the high accretion-rate solution ($\mdot \ga 5000$).
In addition, characteristic features of the line, e.g. a double-peaked spectral shape with a stronger blue peak,
as expected for optically thick spherically collapsing gas clouds \citep{2006ApJ...649...14D,2006ApJ...649...37D}
can serve as useful diagnostics to distinguish hyper-Eddington BHs from other objects.

As we explained, high-energy photons from the central region are hidden by the photon trapping effect.
Thus, the hyper-Eddington BH would not be bright in the X-ray bands.
This fact is consistent with observations that there are only a few detections of 
X-ray emission from LAEs so far\footnote{Local star-forming galaxies emit strong X-rays due to stellar activity
(e.g. high-mass X-ray binaries).
Using an empirical relation between the X-ray luminosity $L_{\rm X}$ and the star-formation rate (SFR),
$L_{\rm X}\sim 10^{39}$ erg s$^{-1}~({\rm SFR}/\msunyr)$ \citep[e.g.][]{2003MNRAS.339..793G}.
Since this value is smaller than the detection limit of current X-ray telescopes, 
the rarity of LAEs associated with X-ray sources is also consistent with a stellar origin.}
\citep[e.g.][]{2004ApJ...608L..21W,2006ApJ...642L..13G,2010ApJ...718...52Z}.
In other words, we propose the possibility that a fraction of LAEs hosts BHs 
which grow at the hyper-Eddington accretion rate.
This would be interesting because most LAEs are not believed to host massive BHs at the center.
Note that even if all high-z LAEs host accreting BHs, the total BH mass density is at most 
$\sim 10^3~\msun~{\rm Mpc}^{-3}$, where we set $M_{\rm BH}=10^6~\msun$
and the number density of LAEs at $z\simeq 6$ is $\sim 10^{-3}~{\rm Mpc}^{-3}$ \citep[e.g.][]{2014ApJ...797...16K},
and is much smaller than that of BHs in the local Universe ($\ga 10^5~\msun$ Mpc$^{-3}$).

Recently, \cite{2015MNRAS.451..400M} found the brightest LAE called CR7.
Spectroscopic observation by \cite{2015ApJ...808..139S} suggests the existence of a Pop III stellar population with
a top-heavy initial mass function.
The Ly$\alpha$ luminosity of CR7 is as large as $9\times 10^{43}$ erg s$^{-1}$, 
which is comparable to the maximum value from the hyper-Eddington accreting BH with $\sim 10^6~\msun$.
Moreover, the non-detection of X-rays from CR7 is also consistent with our proposal that 
X-ray emission is reduced by photon trapping effect.
However, we note that the hyper-Eddington accreting gas could not produce the strong HeII-line of CR7,
because the gas temperature at the photosphere is only $\simeq 10^4~\K$.
Thus, the observed HeII luminosity ($\simeq 2\times 10^{43}$ erg s$^{-1}$)
needs to be explained by massive Pop III stars and/or accreting BHs at a moderate accretion rate 
\citep{Johnson+11,2015MNRAS.453.2465P,2015arXiv151001733A}.

\subsubsection{Luminous infrared galaxies}

Luminous infrared galaxies (LIRGs) are candidates for hosting buried AGN at low redshift.
The typical infrared luminosity is larger than $\sim 10^{11}~\lsun$ \citep{1996ARA&A..34..749S}.
Among them, extremely bright sources called ultra-luminous infrared galaxies (ULIRGs) have an
infrared luminosity of $10^{12}\leq L_{\rm IR}/\lsun \leq 10^{13}$ \citep{1988ApJ...325...74S}.
The mid-infrared observations are useful for diagnosing whether U/LIRGs host AGNs,
because the mid-IR radiation would be emitted from hot circumnuclear dust.
From the spectroscopic data of nearby U/LIRGs, buried AGNs contribute several percent of 
the bolometric luminosity \citep{2012ApJ...744....2A,2014ApJ...794..139I}.
Assuming the same ratio for U/LIRG in the high-$z$ Universe, the luminosity from the accreting BH 
is $\simeq 10^{43-45}$ erg s$^{-1}$.
The corresponding BH mass is estimated as $M_{\rm BH}\simeq 10^{5-7}~\msun$
if the AGN luminosity is near the Eddington luminosity.

We suppose that the accreting gas is polluted by dust grains slightly ($Z\la 10^{-3}~\zsun$)
so that radiation processes (cooling, heating and radiation pressure) involving metals
do not significantly affect gas dynamics in the interior.
Then, we estimate the distribution of dust grains and dust temperature in the outer observable parts of the accretion flow.
From our simulations, the photospheric luminosity is $L_{\rm Edd}$ and 
the effective temperature is $T_{\rm ph}\simeq 10^4~\K$.
We assume that the dust temperature is determined by the balance between 
absorption and emission of thermal radiation,
\begin{align}
T_{\rm dust}(r)&\approx \left(\frac{L_{\rm Edd}}{4\pi \sigma_{\rm SB}r^2}\right)^{1/4},\nonumber\\
&\simeq 2000~\K~M_{\rm BH,4}^{1/4}
\left(\frac{r}{10^{16}~{\rm cm}}\right)^{-1/2}.
\end{align}
Since the dust grains evaporate above the dust sublimation temperature $\approx 2000~\K$,
the dust destruction front is located at 
$R_{\rm dust}\simeq 10^{16}M_{\rm BH,4}^{1/2}(T_{\rm dust}/2000~\K)^{-2}$ cm.

Following \cite{2007ApJ...665.1038C}, we adopt a simple model of dust opacity against 
optical and infrared radiation
\begin{equation}
\kappa_{\rm dust}^{\rm opt}=\frac{300}{1+(T/10^4~\K)}\left(\frac{Z}{\zsun}\right)
{\rm cm}^2~{\rm g}^{-1},
\end{equation}
and $\kappa_{\rm dust}^{\rm IR}=\kappa_{\rm dust}^{\rm opt}/150$, 
where the gas temperature is set to $T=10^4~\K$.
Then, the optical depth of the dust grains against the optical radiation is 
\begin{align}
\tau_{\rm dust}^{\rm opt}&=\int ^\infty_{R_{\rm dust}}\rho \kappa_{\rm dust}^{\rm opt}dr
=\frac{\mdot \kappa_{\rm dust}^{\rm opt}}{\kappa_{\rm es}c}\sqrt{\frac{2GM_{\rm BH}}{R_{\rm dust}}},\nonumber\\
&\simeq 0.24~\mdot_3M_{\rm BH,4}^{1/4}
\left(\frac{Z}{10^{-3}~\zsun}\right) \left(\frac{T_{\rm dust}}{2000~\K}\right).
\end{align}
Since $\tau_{\rm dust}^{\rm opt}$ becomes as large as unity for $\mdot \simeq 5000$, 
most of the radiation from the photosphere is absorbed by dust at $\sim R_{\rm dust}$
and the energy is re-emitted as IR radiation.
The accretion flow is optically thin against the self-absorption of the IR radiation 
as long as $\tau_{\rm dust}^{\rm IR}=\tau_{\rm dust}^{\rm opt}/150\la 1$.
The conditions are written as 
\begin{equation}
M_{\rm BH}\la 4\times 10^5~\msun~n_{\infty,5}^{-4/5}
\left(\frac{Z}{10^{-3}~\zsun}\right)^{-4/5}\left(\frac{T_{\rm dust}}{2000~\K}\right)^{-4/5},
\end{equation}
where we set $\dot{M}=\dot{M}_{\rm B}$.
Thermal radiation from the hottest dust with $T_{\rm dust}\simeq 2000~\K$, 
whose peak wavelength is $\simeq 1.5~\mu$m, can escape from the accreting gas.
On the other hand, for $M_{\rm BH}> 4\times 10^5~\msun$, the re-emitted IR radiation is self-absorbed by dust.
As a result, the dust photosphere defined by $\tau_{\rm dust}^{\rm IR}=1$ moves outwards and 
the dust temperature at the photosphere decreases.
To discuss the observational signature more precisely, we further need to include 
the detailed treatment of dust properties and multi-frequency radiation transfer. 
\citep{2014ApJ...789...78H}, and consider the effects of dust on the gas properties with a hyper-Eddington accretion rate.
We leave this to future work (but note that \citealt{2012MNRAS.425.2892W} discussed the case with $\mdot \ga 1$).


\section{Discussion}
\label{sec:discussion}

\subsection{Multi-dimension effects}
\label{sec:multi}

In this paper, we focus on spherically symmetric accretion onto a BH,
assuming a low angular momentum so that the centrifugal radius is smaller than the trapping radius.
However, even in such a case, collimated outflows and jets from the inner accretion disk in the vertical direction
could potentially break the hyper-Eddington accretion solution
\citep[e.g.][]{2002ApJ...568L..97B,2005ApJ...628..368O,2009PASJ...61L...7O,
2011MNRAS.413.1623D,2013ApJ...767..105L,2014ApJ...796..106J,2014MNRAS.441.3177M,
2015MNRAS.447...49S,2015PASJ...67...60T,2015MNRAS.454L...6M,2016MNRAS.455.1211K}.
The outflows and jets possibly penetrate through the inflow in a narrow conical region as in gamma-ray bursts
\citep[e.g.][]{2003MNRAS.345..575M,2011ApJ...740..100B,2011ApJ...726..107S,2015ApJ...810...64M}
and carry away a large fraction of the gas and radiation energy.
To address whether the radiation or the outflows from the inner disk $\la 100~R_{\rm Sch}$
can reach this trapping radius $\sim 5000~R_{\rm Sch}$ and the bulk properties of the inflow would be affected,
we need to conduct multi-dimensional radiation hydrodynamical simulations with such a large simulation box.

If a steady accretion flow through the disk and narrow conical outflows/jets can coexist, 
the system would be a bright X-ray source with 
$L_{\rm X}\sim 5000~\eta_{\rm X}L_{\rm Edd} \sim 7\times 10^{43}~(M_{\rm BH}/10^5~\msun)(\eta_{\rm X}/10^{-3})$ erg s$^{-1}$
only into the narrow regions.
Assuming that the source exists at $z=6~(10)$, the corresponding X-ray flux is 
$\sim 2\times 10^{-16}~(5\times 10^{-17})$ erg s$^{-1}$ cm$^{-2}$,
which is above the flux threshold of Chandra Deep Field South (CDF-S) with a 4 Msecond exposure
($\simeq 1.5\times 10^{-17}$ erg s$^{-1}$ cm$^{-2}$ in $0.5-2$ keV).
In the CDF-S survey, three X-ray sources ($N_{\rm X}=3$) are known at $6<z<10$
near the observation limit within a sky region of $\sim 170$ arcmin$^2$ \citep{2015A&A...578A..83G}.
Therefore, if the X-ray sources are counterparts of hyper-Eddington accretion BHs, 
the intrinsic number density is given by
$\sim (N_{\rm X}/\Omega_{\rm CDF} V) (t_{\rm obs}/t_{\rm acc}) (2\theta^2)^{-1}\sim 0.5$ Mpc$^{-3}$,
where $\Omega_{\rm CDF}=1.1\times 10^{-6}$ is the sky fraction observed by the CDF-S, 
$V=1.3\times 10^{12}$ Mpc$^3$ is the comoving volume of the Universe between $z=6$ and $10$,
$t_{\rm obs}=464$ Myr is the difference of the cosmic age between $z=6$ and $10$ \citep{2015MNRAS.454.3771P}, 
$t_{\rm acc}=M_{\rm BH}/(5000~\dot{M}_{\rm Edd})\simeq 0.09$ Myr is the accretion time in the hyper-Eddington phase,
and $\theta \simeq 6^\circ$ is the opening angle of the collimated outflows/jets \citep{2007A&A...466..127G}.
Since the hyper-Eddington phase is maintained until the BH grows up to $\sim 10^5~\msun$,
the BH mass density is estimated as $\sim 5\times 10^4~\msun$ Mpc$^{-3}$.
This value is consistent with a constraint from observations of X-ray background, 
$\la 2\times 10^{4}~\msun$ Mpc$^{-3}$ \citep{2012A&A...545L...6S}, as well as with 
the local SMBH mass density \citep{1982MNRAS.200..115S,2002MNRAS.335..965Y}.
Future work is needed to investigate the observability of the objects in detail.

The Rayleigh-Taylor (RT) instability (e.g. \citealt{2013ApJ...763..102J}; 
see also thermal instabilities in, e.g. \citealt{2011MNRAS.418..591B})
would break spherical symmetry of the accretion flow
but would potentially help the accretion.
The instability develops at the boundary of the H$_{\rm II}$ region surrounded by neutral gas
when the gas accretion is suppressed by radiative heating ($\mdot \la 10$).
For the case with burst-like accretion ($M_{\rm BH}=10^3~\msun$; green curve in Fig. \ref{fig:t_Mdot_high}),
the RT instability would occur but not affect the mean accretion rate 
as \cite{2011ApJ...739....2P} found by their 2D simulations.
On the other hand, for $10^3<M_{\rm BH}<10^4~\msun$ (blue and magenta curves in Fig. \ref{fig:t_Mdot_high}), 
the accretion behavior is marginal between the burst-like and the hyper-Eddington accretion.
According to numerical experiments by \cite{2014MNRAS.437.2856P}, 
the RT instability is suppressed if the expansion of the H$_{\rm II}$ region stops,
though the gas piles up just outside the H$_{\rm II}$ region and density inversion occurs.
Thus, the critical accretion rate for hyper-Eddington accretion would not change from our result 
of $\mdot\simeq 5000$ significantly.

\subsection{Supersonic solution without an embedded accretion disk}

In \S\ref{sec:inner}, we determine the physically consistent value of $L_{\rm ph}$ for the hyper-Eddington 
accretion solution with $\mdot\sim 8000$ so that it would link to a thick-disk solution obtained 
from numerical simulations \citep{2015MNRAS.447...49S}.
The hyper-Eddington accretion solution has a subsonic velocity well within $R_{\rm tr}$ 
due to the radiation force (see Fig. \ref{fig:r_L_v}).
Thus, an accretion disk at the bottom is needed to realize the transsonic solution.
We now discuss two possible cases to realize the spherically symmetric hyper-Eddington accretion flow 
on to the central BH keeping a supersonic velocity, without assuming an embedded accretion disk.

The first case is the very low-luminosity solution with $\sim 10^{-5}~L_{\rm Edd}$ as we mentioned in \S\ref{sec:inner}. 
Since the comoving luminosity does not reach $L_{\rm Edd}$ until the gas reaches the BH horizon, 
the accretion flow keeps a supersonic velocity
\citep[e.g.][]{1980ApJ...240..235G,1982MNRAS.199..833F,1984ApJ...284..394V,
1986ApJ...308..755B,1991ApJ...376..234H}.

The second case is shock formation outside the trapping radius.
When the gas experiences deceleration within the trapping radius as shown in Fig. 
\ref{fig:r_L_v}, 
the radiation-dominated gas cannot re-accelerate.
On the other hand, if the gas forms a shock and decelerates well outside the trapping radius, 
where the post-shock gas is not radiation-dominated, the gas inflow can pass the second sonic point again
if the effective adiabatic index decreases from $\approx 5/3$ to $4/3$ in the shocked gas \citep{1985ApJ...288..428C}.
Dissipation due to magnetic fields and Compton heating could result in such shocks.
However, since our simulations include neither, we do not find any transsonic accretion flow with shocks.
We plan to discuss the possibility of self-consistent solutions with shocks in future work
\citep{1985ApJ...288..428C}.

\section{Summary and conclusion}

In this paper, we have studied spherically symmetric accretion flows 
with a very high rate onto massive seed BHs ($10^2 \la M_{\rm BH} \la 10^6~\msun$)
embedded in dense gas clouds with a low abundance of metals. 
We performed one-dimensional hydrodynamical simulations 
which include multi-frequency radiation transfer and non-equilibrium primordial chemistry. 
We find that the condition required to realize steady hyper--Eddington accretion
which overcomes radiative feedback is
\begin{equation}
\left(\frac{n_{\infty}}{10^5~\cc}\right) \ga 
\left(\frac{M_{\rm BH}}{10^4~\msun}\right)^{-1}
\left( \frac{T_{\infty}}{10^4~\K} \right)^{3/2},\nonumber
\end{equation}
where $n_{\infty}$ and $T_\infty$ are the density and temperature of the ambient gas around the BH.
This inequality can be rewritten as 
\begin{equation}
\dot{M}_{\rm B}\ga  5000~\dot{M}_{\rm Edd} = 5000~L_{\rm Edd}/c^2.
\end{equation}
Below the critical value, the accretion becomes episodic due to radiative feedback
and the average rate is limited to $\la 10~L_{\rm Edd}/c^2$. 
Again, we emphasize that energy feedback due to radiative heating dominates over 
momentum feedback due to the radiation pressure force.
We give a simple physical explanation of the transition to hyper-Eddington accretion,
by comparing the Bondi radius and the size of the H$_{\rm II}$ region.

Next, for the hyper-Eddington accretion case, we find a self-consistent steady solution of the accretion flow
from the Bondi radius inward extending inside the trapping radius.
The solution consists of two parts: a radiation-dominated central core, where photon trapping due to 
electron scattering is effective, and an accreting envelope following a Bondi profile with $T\simeq 8000~\K$.
When the emergent luminosity is limited to the Eddington luminosity because of photon trapping, 
radiation from the central region does not affect the gas dynamics at larger scales.
Moreover, we argued that the hyper-Eddington accretion solution with $\mdot \ga 5000$ and $l\simeq 1$ 
could link to recently computed hyper-Eddington accretion disk solutions in the central region ($\la 10^2~R_{\rm Sch}$).

We apply our simulation results to the growth of massive seed BHs in protogalaxies with virial temperature of 
$T_{\rm vir}\ga 10^4~\K$.
Once the seed BH forms at the center of the galaxy, 
it can grow up to $\sim 10^5~(T_{\rm vir}/10^4~\K)~\msun$ via rapid gas accretion,
independent of the BH initial mass, i.e. Pop III remnant BHs or direct collapse BHs.
The massive BH with $\ga 10^5~\msun$ formed by the hyper-Eddington accretion
could grow in mass to explain the $z>6$ quasar BHs by subsequent moderate sub-Eddington growth.
Finally, we discuss observational signatures of rapidly accreting BHs with/without allowance for dust. 
We suggest that these systems could explain Ly$\alpha$ emitters without X-rays and luminous infrared sources 
with hot dust emission, respectively.

\section*{Acknowledgements}
We thank Aleksander S{\c a}dowski for providing his simulation data for comparisons 
with our spherical symmetric accretion flows, and 
Martin Rees, Kazumi Kashiyama, Ken Ohsuga, Kengo Tomida, Andrea Ferrara, Massimo Ricotti,
Jarrett Johnson, Tal Alexander, KwangHo Park, Muhammad Latif and Fabio Pacucci for fruitful discussions.
This work is partially supported by the Grants-in-Aid by the Ministry of Education, 
Culture and Science of Japan, and by the Simons Foundation through 
the Simons Society of Fellows (KI), and by NASA grant NNX11AE05G and NNX15AB19G (ZH).

\appendix
\section{cooling function and opacity}

In our paper, we do not solve radiation transfer equations for photons with $h\nu<13.6$ eV.
Instead, we treat them by using fitting formulae for cooling rates and opacities.
In the inner-region simulations, the main cooling processes are 
Ly$\alpha$ line emission ($10.2$ eV), two-photon emission ($\la 10.2$ eV), 
and free-bound emission of H$^-$ ($0.76-13.6$ eV).
In addition, bound-free absorption of H$^-$, electron scattering and hydrogen Rayleigh scattering 
work as opacity sources.

\subsection{atomic hydrogen (two levels)}
We solve two levels of atomic hydrogen to estimate the cooling rates due to 
Ly$\alpha$ emission ($2P\rightarrow 1S$) and two-photon emission ($2S\rightarrow 1S$).
Our treatment is based on that in \cite{O01} and \cite{2006ApJ...652..902S}.
Both the cooling rates are given by
\begin{equation}
\Lambda_{{\rm Ly}\alpha}=h\nu_{21}\beta_{\rm esc,21}A_{21}n_{\rm 2p}
\end{equation}
\begin{equation}
\Lambda_{{\rm 2ph}}=h\nu_{21}\beta_{\rm esc,2ph}A_{\rm 2ph}n_{\rm 2s},
\end{equation}
where $h\nu_{21}=10.2$ eV, $\beta_{\rm esc,21}$ is the escape fraction of Ly$\alpha$ photons,
$\beta_{\rm esc,2ph}=\exp(-\tau_{\rm eff})$ is the escape fraction of two-photon emissions (see Appendix A3),
$A_{21}$ ($A_{\rm 2ph}$) is the Einstein A-coefficient for Ly$\alpha$ (two-photon) emission, and
$n_{\rm 2p(2s)}$ is the number density of atomic hydrogen which occupies $2P$ ($2S$) state.
The level population is determined by the equation of the detailed balance between each hydrogen levels.
Here, we consider only three states ($1S$, $2S$, and $2P$ state) and neglect higher level populations for simplicity.
The detailed balance is written by
\begin{equation}
n_{2}R_{21}=n_1R_{12},
\end{equation}
where $R_{21}$ ($R_{12}$) is the transition rate from the excited (ground) state to the ground (excited) state,
\begin{align}
R_{21}&=A_{21}\beta_{\rm esc,21} + A_{\rm 2ph} +C_{21},\\
R_{12}&=C_{12},
\end{align}
where we neglect the Einstein B coefficient due to thermal radiation from the photosphere
because the optical depth to the Ly$\alpha$ photons is so large around the photosphere.
The relative population within the first excited level ($2S$ and $2P$ states) is given by
\begin{equation}
r_{\rm 2s2p}=\frac{n_{\rm 2s}}{n_{\rm 2p}}=\frac{g_{\rm 2s}}{g_{\rm 2p}}\left(\frac{C_{\rm 2s2p}}{C_{\rm 2s2p}+A_{\rm 2s1s}}\right),
\end{equation}
where $g_{\rm 2s}=2$ and $g_{\rm 2p}=6$, the radiative transition rate by two-photon emissions is 
$A_{\rm 2s1s}=8.23$ s$^{-1}$, and the collisional transition rate between these levels is $C_{\rm 2s2p}(n,T,x_{\rm e})$ \citep{O01}.
From the relative abundance between $2S$ and $2P$ state, 
\begin{align}
A_{21}&=\frac{n_{\rm 2p}}{n_2}A_{2p1s}=\frac{1}{1+r_{\rm 2s2p}}A_{\rm 2p1s},\\
A_{\rm 2ph}&=\frac{n_{\rm 2s}}{n_2}A_{2s1s}=\frac{r_{\rm 2s2p}}{1+r_{\rm 2s2p}}A_{\rm 2s1s},
\end{align}
where $A_{\rm 2p1s}=6.27\times 10^{8}$ s$^{-1}$.
The collisional de-excited rate $C_{21}$ is given by
\begin{equation}
C_{21}=\gamma_{21}({\rm e})n({\rm e}) + \gamma_{21}({\rm H})n({\rm H}),
\end{equation}
and the collisional excited rate is given by
\begin{equation}
C_{12}=C_{21}(g_{2}/g_{1})\exp(-h\nu_{21}/k_{\rm B}T),
\end{equation}
so that the level population approaches the Boltzmann distribution.

Next, we estimate the escape probability $\beta_{\rm esc,21}$ as
\begin{equation}
\beta_{\rm esc,21}=\frac{p_{21}^{N_{\rm reflect}}}{1+N_{\rm reflect}},
\end{equation}
where $p_{21}$ is the probability for an absorbed Ly$\alpha$ photon to re-emerge as a Ly$\alpha$ photon
with neither collisional de-excitation nor two-photon emission, and $N_{\rm reflect}$ is the number of scatterings
that an average Ly$\alpha$ photon experiences before escape.
The re-emerge probability is given by
\begin{equation}
p_{21}=\frac{A_{21}}{A_{21}+A_{\rm 2ph}+C_{21}}.
\end{equation}
The number of scatterings is estimated by $N_{\rm reflect}\approx 15~(\tau_{21}/10^{5.5})^{1/3}$
\citep{1975ApJ...201..350A,2008MNRAS.391..457D}
where $\tau_{21}=\sigma_{\rm Ly\alpha} n r$ 
($\sigma_{\rm Ly\alpha}=5.9\times 10^{-14}T_4^{-1/2}$ cm$^2$).
Note that the above estimate neglects the effect of gas dynamics.
Following \cite{2006ApJ...652..902S}, we consider the dynamical effect, that is,
when the timescale within which the Ly$\alpha$ photons are trapped, $t_{\rm Ly\alpha}(=N_{\rm reflect}r/c)$,
is longer than the dynamical timescale $t_{\rm dyn}(=r/|v|$), 
the cooling efficiency decreases rapidly.
We use the model of the escape probability shown in \cite{2006ApJ...652..902S}, 
\begin{equation}
\beta_{\rm esc,21}\rightarrow \beta_{\rm esc,21}\exp\left(-3 \frac{t_{\rm Ly\alpha}}{t_{\rm dyn}}\right).
\end{equation}

For convenience, we define the Ly$\alpha$ trapping radius $R_{\rm Ly\alpha}$, 
within which two-photon emission dominates Ly$\alpha$ emission 
(i.e. $\Lambda_{\rm 2ph}>\Lambda_{\rm Ly\alpha}$) because of efficient Ly$\alpha$ trapping.
We fit the position of Ly$\alpha$ trapping radius as a function of the BH mass and ambient density, 
respectively
\begin{equation}
R_{\rm Ly\alpha}\approx 7\times 10^{14}~{\rm cm}~M_{\rm BH,4}^{0.55}n_{\infty,5}^{0.47},
\label{eq:lya_tr}
\end{equation}
where the isothermal Bondi profile with $T=8000~\K$ and a constant electron fraction ($x_{\rm e}=10^{-4}$) are assumed. 
In fact, this estimate agrees with the simulation result ($\simeq 5\times 10^{14}$ cm) for 
$M_{\rm BH}=10^4~\msun$ and $n_\infty=10^5~\cc$.

\subsection{continuum cooling (H$^-$ free-bound)}
At high density regime in the inner-region simulation, 
H$^-$ free-bound emission (${\rm H} + {\rm e}^- \rightarrow {\rm H}^- + \gamma$) 
becomes an important cooling process as well as two-photon emission.
We estimate the emissivity of the radiative association in what follows 
\citep{1984oup..book.....M, O01}.

The emissivity of continuum radiation due to H$^-$ free-bound transition is given by
\begin{equation}
\eta_\nu ^{\rm H^-}=\frac{2h\nu^3}{c^2} \frac{z_{\rm H^-}}{z_{\rm H}z_{\rm e}}
\left(\frac{h^2}{2\pi m_{\rm e} k_{\rm B}T}\right)^{3/2} 
\sigma_{\nu}^{\rm H^-}
e^{- \frac{h(\nu - \nu_0)}{k_{\rm B}T}}n({\rm H})n({\rm e}),
\end{equation}
where $z_i$ is the partition function of $i$-th species ($z_{\rm H^-}=1$, $z_{\rm e}=2$, and 
$z_{\rm H}=\Sigma_{\rm n=1}^{\infty} ~ g_{\rm n} \exp(-E_n/k_{\rm B}T)$, where $g_{\rm n}=2n^2$ 
and $E_{\rm n}=13.6/n^2~{\rm eV}$), 
$m_{\rm e}$ is the electron mass, 
$h\nu_0=0.76$ eV is the dissociation energy of H$^-$, and $\sigma _\nu^{\rm H^-}$ is 
the cross section of H$^-$ photodissociation (${\rm H}^- + \gamma \rightarrow {\rm H} + {\rm e}^-$) \citep{1988A&A...193..189J}. 
From these expressions, the optically-thin cooling rate (in units of erg s$^{-1}$ cm$^{-3}$) 
is calculated by integration over the frequency
($\Lambda_{\rm H^-}^{\rm thin}=4\pi \int _{\nu_0}^{13.6{\rm eV}/h}\eta_\nu d\nu$) as 
\begin{align}
\Lambda_{\rm H^-}^{\rm thin}=&~10^{-27}~T_3~
(1.4924 + 0.07815~T_3 + 0.0063~T_3^2)\nonumber \\
&\times (1.0 - 0.1535~T_3^{0.5})~n({\rm H})n({\rm e}),
\end{align}
where $T_3=T/10^3~\K$.
For higher density, the gas becomes opaque against continuum radiation because of absorptions and scatterings.
We here reduce the cooling rate using the effective optical depth $\tau_{\rm eff}$ (see Appendix A3) as
\begin{equation}
\Lambda_{\rm H^-}=\frac{\Lambda_{\rm H^-}^{\rm thin}}{1+\tau_{\rm eff}^2},
\end{equation}
so that the cooling rate approaches to the value in diffusion approximation for a large optical depth.

\subsection{opacities}
We consider several opacity sources to estimate the cooling efficiency 
and the position of the photosphere.
In our simulation, H$^-$ bound-free transition, H free-free transition, 
H Rayleigh scattering and electron scattering 
are included as the absorption and scattering processes.
To capture the effect of non-equilibrium chemistry, we make fitting formulae of 
the Rosseland mean opacity due to H$^-$ bound-free transition and H Rayleigh scattering.
For other processes, we adopt the expressions in \cite{1979rpa..book.....R}.

The opacity of H$^-$ bound-free transition is given by
\begin{equation}
\kappa_\nu^{\rm H^-} = \sigma_\nu^{\rm H^-} n({\rm H^-})\left\{1-e^{-h\nu/k_{\rm B}T}
\left[\frac{K(T)}{K_{\rm eq}(T)}\right] \right\},
\label{eq:kappa_Hm}
\end{equation}
where $K(T)=n({\rm H})n({\rm e})/n({\rm H^-})$ and 
\begin{align}
K_{\rm eq}(T)&=\left[\frac{n({\rm H})n({\rm e})}{n({\rm H^-})}\right]_{\rm eq}\nonumber\\
&=\frac{z_{\rm H}z_{\rm e}}{z_{\rm H^-}}\left(\frac{2\pi m_{\rm e}k_{\rm B}T}{h^2}\right)^{3/2}e^{-h\nu_0/k_{\rm B}T}.
\end{align}
We also solve the chemical reaction networks which relates to H$^-$, where the following four are considered:
\begin{align}
{\rm H}+{\rm e^-}&\rightarrow {\rm H}^-+\gamma,\label{eq:rec1}\\
{\rm H}^-+{\rm H}&\rightarrow {\rm H}_2+{\rm e}^-,\label{eq:rec2}\\
{\rm H}^-+{\rm e}^-&\leftrightarrow {\rm H}+2{\rm e}^-\label{eq:rec34}.
\end{align}
The first two reactions are known as H$_2$ formation processes in primordial gas clouds \citep[e.g.][]{1969PThPh..42..219M}.
The last two reactions are necessary to achieve the chemical-equilibrium state at a high-density regime \citep{IOT14}.
The rate coefficients ($k_i,~i=1-4$) are given by \cite{2008MNRAS.388.1627G} (\ref{eq:rec1}), 
\cite{2011ApJS..193....7C} (\ref{eq:rec2}), and \cite{1991ApJS...76..759L} (\ref{eq:rec34}), respectively.
Since these reactions occur faster than those which relate to other species,
we assume $dx_{\rm H^-}/dt \approx 0$ and thus obtain
\begin{equation}
n({\rm H^-})=\frac{k_1 + k_4 n}{k_2n({\rm H}) + k_3 n({\rm e})}n({\rm H})n({\rm e}).
\end{equation}
For typical values of density and temperature in the outer-region simulations,
$0<K(T)/K_{\rm eq}(T)<1$.
Then, the second term of Eq. (\ref{eq:kappa_Hm}) due to the stimulated emission is negligible.
For simplicity, we here calculate the Rosseland-mean opacity from Eq. (\ref{eq:kappa_Hm}) 
by setting $K(T)=K_{\rm eq}(T)$ so that the value is correct in the chemical equilibrium state.
The fitting formula of the H$^-$ bound-free opacity is 
\begin{align}
\rho \kappa_{\rm H^-}&=10^{-18}~n({\rm H^-})\\
&\times \left\{ \begin{array}{ll}
(32.0781 - 248.509~T_4 + 1205.57~T_4^2 \\
- 2343.75~T_4^3 + 2004.93~T_4^4 - 636.453~T_4^5),\\
\vspace{-1mm}\\
(52.3537 - 73.3006~T_4 + 49.7537~T_4^2 \\
- 17.5565~T_4^3 + 3.25622~T_4^4 - 0.246158~T_4^5),\\
\end{array} \right.\nonumber
\end{align}
where the top (bottom) formula is for $T_4=T/10^4~\K \leq (>)1$.

The opacity of H Rayleigh scattering is given by 
$\kappa_\nu^{\rm Ray}=\sigma_\nu^{\rm Ray}n({\rm H})$
\citep{1970acpc.book.....K}.
The fitting formula of the Rosseland mean opacity is expressed as 
\begin{align}
\rho \kappa_{\rm Ray}
&=n({\rm H})~
\frac{3.814\times 10^{-33}~T_3^{4.25}}{(1 + 0.139~T_3^{1.204})}
\exp\left(\frac{8.761}{T_3}\right).
\end{align}

Finally, we define the absorption and scattering opacity for low-energy photons ($0.76 \leq h\nu\leq 13.6$ eV)
as the sum of all the opacities,
\begin{align}
\kappa_{\rm abs}&=\kappa_{\rm H^-}+\kappa_{\rm H,ff},\nonumber\\
\kappa_{\rm scat}&=\kappa_{\rm Ray}+\kappa_{\rm es},
\end{align}
where $\kappa_{\rm H,ff}$ and $\kappa_{\rm es}$ are the opacity of 
H free-free transition and electron scattering, respectively.
Then, we estimate the effective optical depth as 
\begin{equation}
\tau_{\rm eff}(r)=\sqrt{3 \tau_{\rm abs} (\tau_{\rm abs}+\tau_{\rm scat})},
\end{equation}
where $\tau_{\rm abs(scat)}=\int_r^\infty \rho \kappa_{\rm abs(scat)}dr$.
We define the photosphere $R_{\rm ph}$ from $\tau_{\rm eff}(R_{\rm ph})=1$.

As we explained in \S\ref{sec:RT_in}, we calculate the opacities assuming chemical equilibrium at the early stage 
of the inner-region simulation ($t\leq 10^8$ s) in order to save computational time.
We adopt the equilibrium opacities for H$^-$ free-bound transition and electron scattering 
in the same way as \cite{2015MNRAS.448..104P},
\begin{equation}
\kappa_{\rm es}^{\rm eq}=\frac{0.35}{1+(T/8000~\K)^{-13}}~{\rm cm^2~g^{-1}},
\label{eq:eq_op1}
\end{equation}
\citep{2008MNRAS.387.1649B} and 
\begin{equation}
\kappa_{\rm H^-}^{\rm eq}=3.5\times 10^{-27}\rho^{1/2}T^{7.7}x_{\rm H}~{\rm cm^2~g^{-1}},
\label{eq:eq_op2}
\end{equation}
\citep{1996ima..book.....C}.

\section{Pre-heating due to Compton radiation}

In our simulations, we have neglected Compton heating by radiation from a hot accretion flow with 
$T_{\rm comp}>10^8~\K$.
As we introduced in \S\ref{sec:intro}, however, Compton heating potentially destroys steady 
and/or self-consistent solutions \citep{1976ApJ...208L..61O,1978ApJ...226.1041C,1985ApJ...288..428C,
1990ApJ...354...64P,1990ApJ...354...83P,1991ApJ...383..250N}.
We discuss whether the hyper-Eddington accretion solution is invalidated by Compton heating.
To address this, we compare the heating timescale $t_{\rm comp}$ and the radiative cooling timescale $t_{\rm cool}$.
The Compton heating rate is given by
\begin{equation}
\Gamma_{\rm comp}=\frac{4k_{\rm B}T_{\rm comp}}{m_{\rm e}c^2}\frac{L}{4\pi r^2}\sigma_{\rm T}n_{\rm e}.
\end{equation}
The ratio of the two timescales ($t_{\rm comp}/t_{\rm cool}=n\Lambda(T)/\Gamma_{\rm comp}$)
at the Bondi radius is given by 
\begin{align}
\frac{t_{\rm comp}}{t_{\rm cool}}
&\simeq \frac{21M_{\rm BH,4}n_{\infty,5}\Lambda_{-23}}{l~T_{\rm comp,8}T_{\infty,4}^2}
\simeq \frac{4\mdot_3 \Lambda_{-23}}{l~T_{\rm comp,8}T_{\infty,4}^{1/2}},
\end{align}
where $\Lambda_{-23}(\equiv \Lambda/(10^{-23}~{\rm erg~cm^3~s^{-1}}))\simeq 1$ corresponds to
the atomic cooling rate for $T\simeq 10^4~\K$ and $T_{\rm comp,8}\equiv T_{\rm comp}/(10^8~\K)$.
In the quiescent phase for the episodic accretion with $\mdot \la 10$ and $l\simeq 1$, 
the density in the H$_{\rm II}$ region decreases.
Then, the accretion flow there is almost adiabatic because of inefficient radiative cooling ($\Lambda \propto n^2$)
and the gas temperature at $\sim R_{\rm Sch}$ becomes very high 
$T_{\rm comp,8}\gg 1$.
Therefore, Compton heating should result in lower accretion rates than our results 
because $t_{\rm comp}/t_{\rm cool}\approx 4\times 10^{-3}\mdot/(l~T_{\rm comp,8})\ll 1$.
This is consistent with previous works (but see also \citealt{2014MNRAS.445.2325P}).

For the hyper-Eddington accretion case, on the other hand, 
the Compton temperature should be smaller than $\sim 10^8~\K$
(see \S\ref{sec:inner}).
This is because the trapping radius is so large that high-energy radiation from the inner hot gas cannot escape.
Moreover, the gas temperature would not be very high even near the BH
because the temperature still only modestly exceeds $10^4~\K$ in the inner region ($\sim 10^5~R_{\rm Sch}$).
Thus, $t_{\rm comp}/t_{\rm cool}\simeq 17(\mdot/5000)/(l~T_{\rm comp,8})\gg 1$
and Compton heating should not affect the hyper-Eddington accretion solution.

{\small
\bibliography{ref.bib}
}
\end{document}